\pdfoutput=1
\documentclass[preprint,12pt]{aastex}

\newcommand{\degree}{\mbox{$^{\circ}$}}
\newcommand{\am}{\mbox{\arcmin}}
\newcommand{\as}{\mbox{\arcsec}}

\newcommand{\lsun}{\mbox{L$_\odot$}}
\newcommand{\msun}{\mbox{M$_\odot$}}

\input{epsf}

\begin{document}

\slugcomment{v3.00; 23 Dec 2012}

\renewcommand\thefootnote
	{\fnsymbol{footnote}}
\title {\bf  A First Look at the Auriga-California Giant Molecular Cloud With {\it Herschel}\footnote{{\it Herschel} is an ESA space observatory with science instruments provided by European-led Principal Investigator consortia and with important participation from NASA.} and the CSO: Census
of the Young Stellar Objects and the Dense Gas}
\author{Paul M. Harvey\altaffilmark{1},
Cassandra Fallscheer\altaffilmark{2},
Adam Ginsburg\altaffilmark{3},
Susan Terebey\altaffilmark{4},
Philippe Andr\'e\altaffilmark{5},
Tyler L. Bourke\altaffilmark{6},
James Di Francesco\altaffilmark{7},
Vera K\"onyves\altaffilmark{5,8},
Brenda C. Matthews\altaffilmark{7},
\& Dawn E. Peterson\altaffilmark{9}
}

\altaffiltext{1}{Astronomy Department, University of Texas at Austin, 1 University Station C1400, Austin, TX 78712-0259, USA;  pmh@astro.as.utexas.edu}
\altaffiltext{2}{Department of Physics and Astronomy, University of Victoria, 3800 Finnerty Road, Victoria, BC V8P 5C2, Canada; Cassandra.Fallscheer@nrc-cnrc.gc.ca}
\altaffiltext{3}{Center for Astrophysics and Space Astronomy, University of Colorado, 389 UCB, Boulder, CO 80309-0389, USA; adam.ginsburg@colorado.edu}
\altaffiltext{4}{Department of Physics and Astronomy PS315, 5151 State University Drive, California State University at Los Angeles, Los Angeles, CA 90032, USA; sterebe@calstatela.edu}
\altaffiltext{5}{Laboratoire AIM, CEA/DSM-CNRS-Universit\'e Paris Diderot, IRFU/Service d'Astrophysique, CEA Saclay, 91191 Gif-sur-Yvette, France; pandre@cea.fr, vera.konyves@cea.fr}
\altaffiltext{6}{Harvard-Smithsonian Center for Astrophysics, 60 Garden Street, Cambridge, MA 02138, USA; tbourke@cfa.harvard.edu}
\altaffiltext{7}{Herzberg Institute of Astrophysics, National Research Council of Canada, 5071 West Saanich Road, Victoria, BC V9E 2E7, Canada; James.DiFrancesco@nrc-cnrc.gc.ca, Brenda.Matthews@nrc-cnrc.gc.ca}
\altaffiltext{8}{Institut d'Astrophysique Spatiale, CNRS/Universit\'e Paris-Sud 11, 91405 Orsay, France}
\altaffiltext{9}{Space Science Institute, 4750 Walnut Street, Suite 205, Boulder, CO 80303, USA; dpeterson@spacescience.org}

\begin{abstract}

We have mapped the Auriga/California molecular cloud with the {\it Herschel} PACS and SPIRE
cameras and the Bolocam 1.1 mm camera on the Caltech Submillimeter Observatory (CSO) with the eventual goal of quantifying the star formation and
cloud structure in this Giant Molecular Cloud
(GMC) that is comparable in size and mass to the Orion GMC, but which appears to be forming far fewer stars.
We have tabulated 60 compact 70/160 \micron\ sources that are likely
pre-main-sequence objects and correlated those
with {\it Spitzer} and WISE mid-IR sources.  At 1.1 mm
we find 18 cold, compact sources and discuss their properties.
The most important result from this part of our study is that we find a modest number of additional compact young objects
beyond those identified at shorter wavelengths with {\it Spitzer}.
We also describe the dust column density and temperature structure
derived from our photometric maps.  The column density peaks at a few $\times\ 10^{22}$ cm$^{-2}$ ($N_{H2}$)
and is distributed in a clear filamentary structure along which nearly all the pre-main-sequence objects are found.  We compare the YSO
surface density to the gas column density and find a strong non-linear correlation between them.  The dust temperature
in the densest parts of the filaments drops to $\sim$ 10K from values $\sim$ 14--15K in the low density parts of the cloud.
We also derive the cumulative mass fraction and probability
density function of material in the cloud which we compare with similar data on other star-forming clouds.

\end{abstract}


\section{Introduction}\label{intro}

The Auriga-California molecular cloud (AMC) is a large region of relatively modest star formation that is part of the Gould Belt.
We have adopted the name ``Auriga-California Molecular Cloud'' since the region is listed as ``Auriga'' in the
{\it Spitzer Space Telescope} \citep{werner04} Legacy Survey by L. Allen, while it has been called the ``California Molecular
Cloud'' by \citet{lada09} based on its proximity to the ``California Nebula''.  The {\it Spitzer} observations of this region are described
by H. Broekhoven-Fiene et al. (2013, in preparation) as part of the large scale {\it Spitzer} ``From Cores to Planet-Forming Disks'' ({\it c2d}) and ``Gould Belt''
programs that were aimed at cataloguing the star formation in the solar neighborhood.  A similar large-scale
mapping program with the {\it Herschel Space Observatory} \citep{pilbrat10}, the ``Herschel Gould Belt Survey'' (KPGT1\_pandre\_01) \citep{andre10}, has been observing most
of the same star-forming regions, but the AMC was not included in the original target list for that program.

The AMC provides an important counterpoint to other star-forming regions in the Gould Belt, particularly the
well-known Orion Molecular Cloud (OMC). As described first by \citet{lada09}, the AMC is at a likely distance of 450 pc (though \citet{wolk10} quote
a slightly larger distance of 510 pc).
This distance is quite
comparable to that of the OMC, and the mass of the AMC estimated by \citet{lada09} is also quite similar, $\sim 10^5$ \msun.
The most massive star that is forming in the AMC, however, is probably the Herbig emission-line star LkH$\alpha$101,
likely an early B star embedded in a cluster of lower
mass young stars \citep{sfhb08,herbig04}.  This situation is in stark contrast to the substantial number of OB stars found in several
tight groupings in the OMC \citep{blaauw64}.  \citet{lada09} investigated the distribution of optical extinction in the AMC and used those
results together with $^{12}$CO maps from \citet{dame01} to conclude that one possibly significant difference between the
AMC and OMC is the much smaller total area exhibiting high optical extinction in the AMC, roughly a factor of 6 smaller area
above $A_K = 1$ mag.

{\it Herschel} observations have
demonstrated probably the best combination of sensitivity and angular resolution to a range of dust column densities in star-forming
regions, as
well as excellent sensitivity to the presence of star formation from the very earliest stages to the so-called Class II objects
with modest circumstellar disks.  We therefore have undertaken a {\it Herschel} imaging survey of a $\sim$ 15 deg$^2$ area of
the AMC to document the full range in evolutionary status of the star formation in this cloud as well as the distribution
and column density of dust as a proxy for the total mass density.  We have supplemented the {\it Herschel} observations with
a 1.1 mm Bolocam map from the Caltech Submillimeter Observatory (CSO) to identify the extremes in cold, dense material.
We describe the observations and data reduction in the following section.  Then in \S\ref{compact} we discuss
our extraction of the compact source component in the 70/160 \micron\ {\it Herschel} data as well as in the
1.1 mm maps and compare our fluxes with those from other measurements. In \S\ref{indiv} we describe several interesting
individual objects.  In \S\ref{diff} we discuss 
the dust column density and temperature maps derived from our {\it Herschel} PACS/SPIRE images and the relationship between this dust emission
and previous observations of dust absorption and gas emission.  We also derive a quantitative correlation between the
gas density and YSO surface density.  Finally in \S\ref{compare} we begin a discussion of the
differences between star formation in the AMC versus that in the OMC, a subject which we will investigate more fully
in future studies.

\section{Observations and Data Reduction}\label{obs}

\subsection{{\it Herschel} Observations}

\begin{figure}
\includegraphics[angle=0,totalheight=100mm]{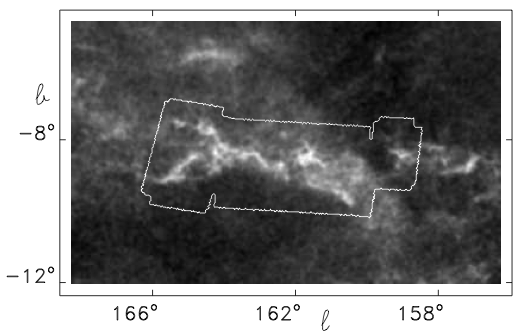}
\caption{\label{aurcover}
Extinction image of \citet{dob05} with the outline of our covered area shown to illustrate that
we have observed most of the high extinction parts of this cloud.  The maximum extinction is A$_v = 4.5$ mag and
the minimum is essentially zero.
The image is oriented in Galactic coordinates and covers an area of 12\degree\ ({\it l}) $\times$ 7.3\degree\ ({\it b}).}
\end{figure}

Our {\it Herschel} program, the ``Auriga-California Molecular Cloud'' (OT1\_pharve01\_3), was designed to use the same observing modes as  comparable parts of the
large-scale Gould Belt program by \citet{andre10}.  For both programs the ``Parallel Mode'' of PACS/SPIRE \citep{spire} was used
to cover the largest possible size region in a reasonable observing time, and a much smaller region was covered
with PACS \citep{pacs10} alone to provide additional sensitivity and wavelength coverage.  The Parallel Mode observations were
done with PACS at 70 \micron\ and 160 \micron, and the SPIRE observations naturally included the three SPIRE photometric
bands, 250 \micron, 350 \micron, and 500 \micron, that are observed simultaneously.  With the PACS-only observations, as for
the larger scale {\it Herschel} Gould Belt program, we used PACS at 100/160 \micron\ with slow scan speed (20\as/s) which essentially
preserves the full diffraction-limited resolution of {\it Herschel}.  These
latter observations were centered on the well-known LkH$\alpha$101 cluster \citep{sfhb08} which includes a
significant fraction of all the obvious star formation in this cloud.  We do not discuss these PACS-only observations further
in this paper, but they will be used in a subsequent study to help address source confusion in the dense central cluser.
The total area covered in Parallel Mode is
18.5 deg$^2$ with 14.5 deg$^2$ covered with overlapping perpendicular scans for good drift cancellation.  
Figure \ref{aurcover} shows the area covered in Parallel Mode overlaid on the extinction map of a much larger portion
of this area discussed by \citet{dob05}.  Our covered area was chosen to include essentially all of the high-extinction
parts of the cloud with the exception of L1441 which is beyond the right (low Galactic longitude) end of our maps.
The PACS-only observations covered 1.4 deg$^2$.  The details of the observations and ObsIDs are listed in Table \ref{obsidtbl}.
The observed Parallel-Mode area was divided into three separate pieces for efficiency in AOR design and observatory scheduling.
The area covered includes nearly all of that observed by the {\it Spitzer} Gould Belt study of the AMC with the exception
of a small separate portion northwest of the end of our maps.

\begin{figure}
\includegraphics[angle=90,totalheight=80mm]{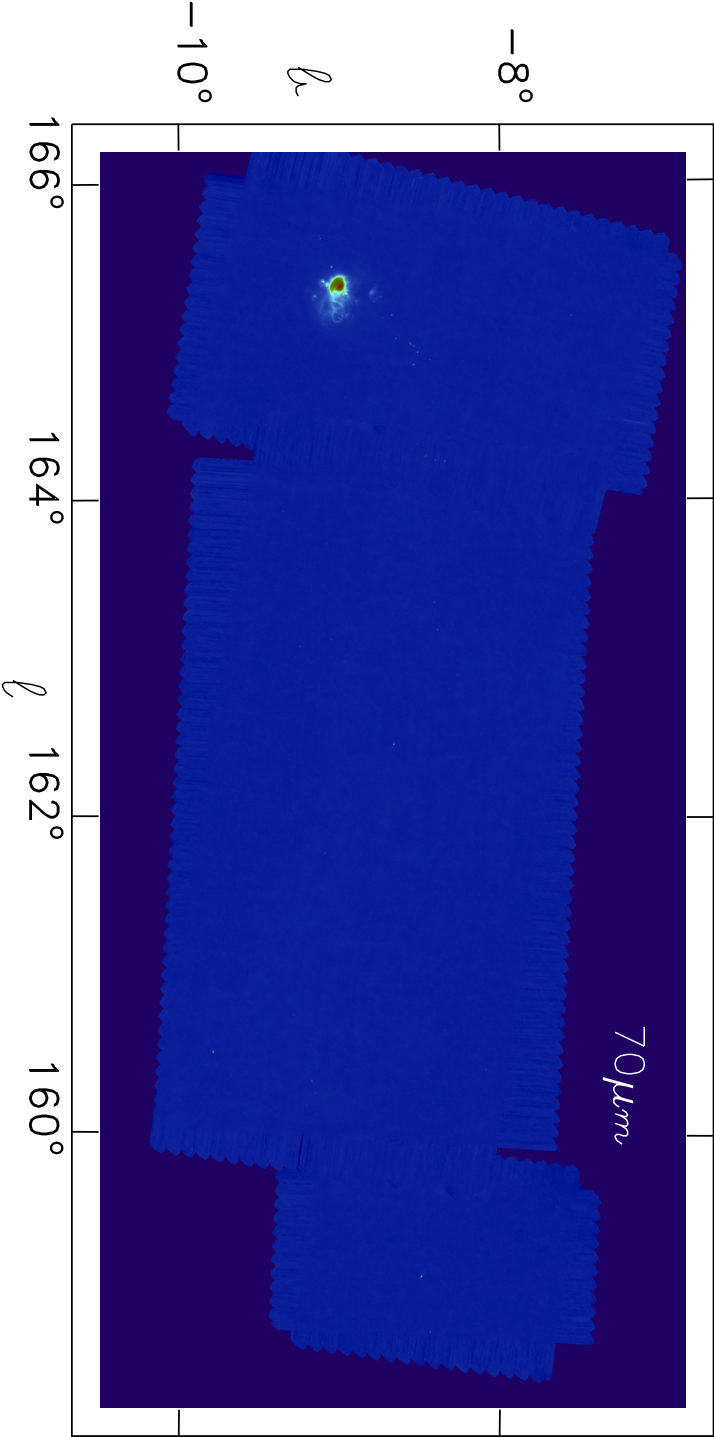}
\figcaption{\label{aur70}
False color image of the 70 \micron\ map derived with Scanamorphos from our Parallel-Mode observations.}
\end{figure}

\begin{figure}
\includegraphics[angle=90,totalheight=80mm]{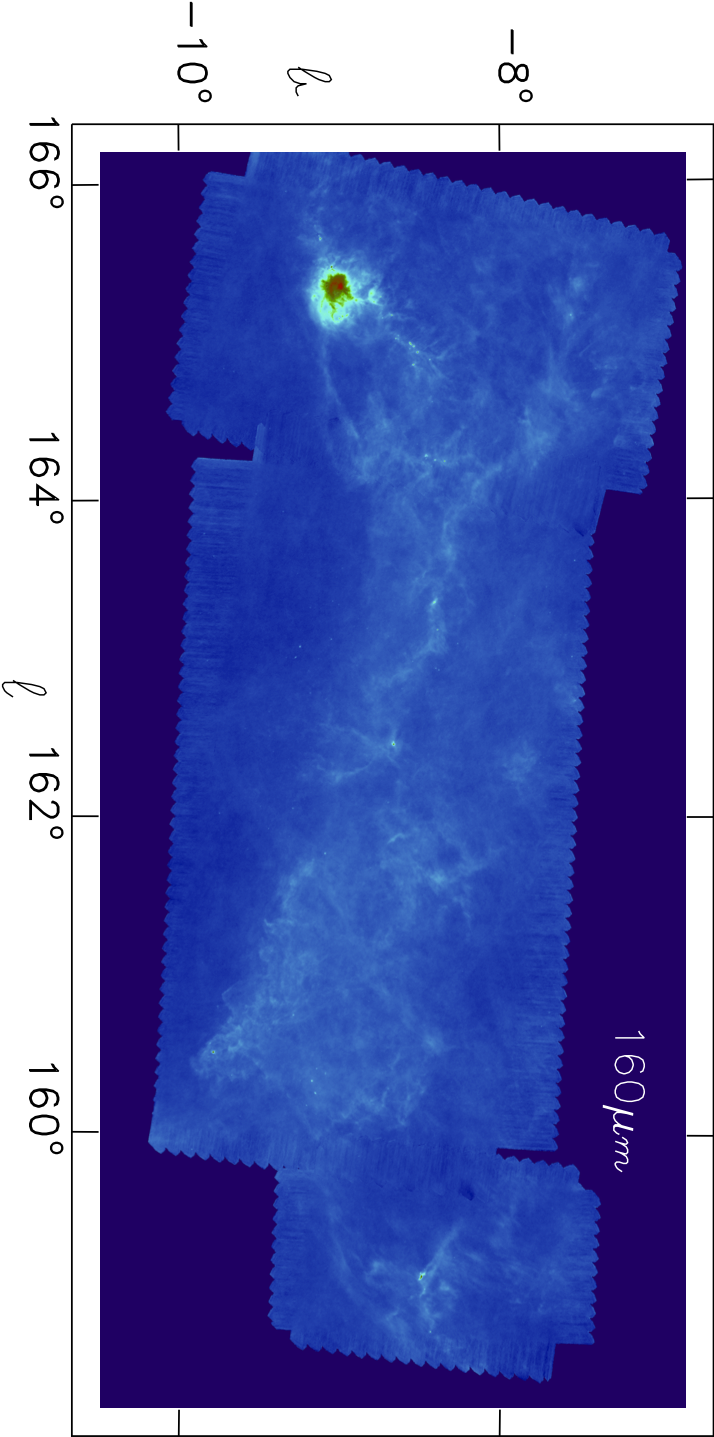}
\figcaption{\label{aur160}
False color image of the 160 \micron\ map derived with Scanamorphos from our Parallel-Mode observations.}
\end{figure}

\begin{figure}
\includegraphics[angle=90,totalheight=80mm]{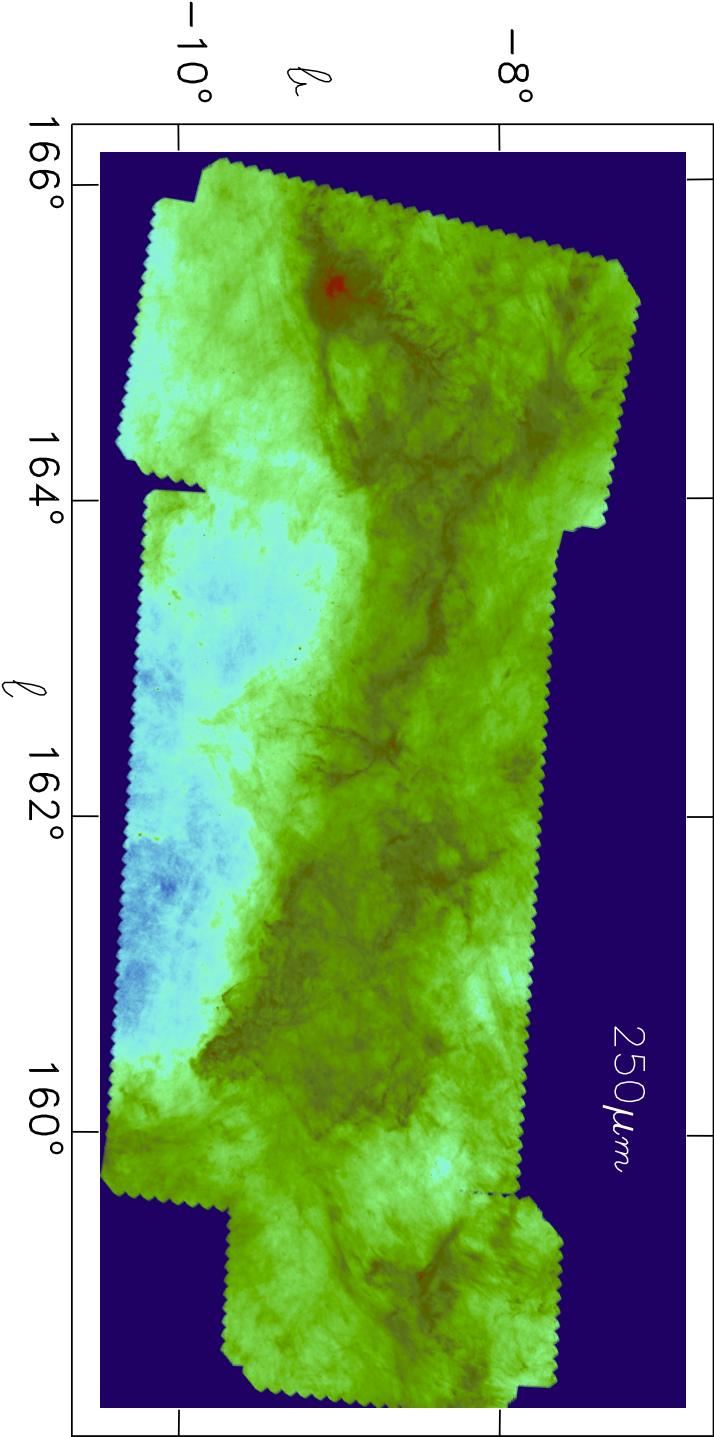}
\figcaption{\label{aur250}
False color image of the 250 \micron\ map derived with Scanamorphos from our Parallel-Mode observations.}
\end{figure}

\begin{figure}
\includegraphics[angle=90,totalheight=80mm]{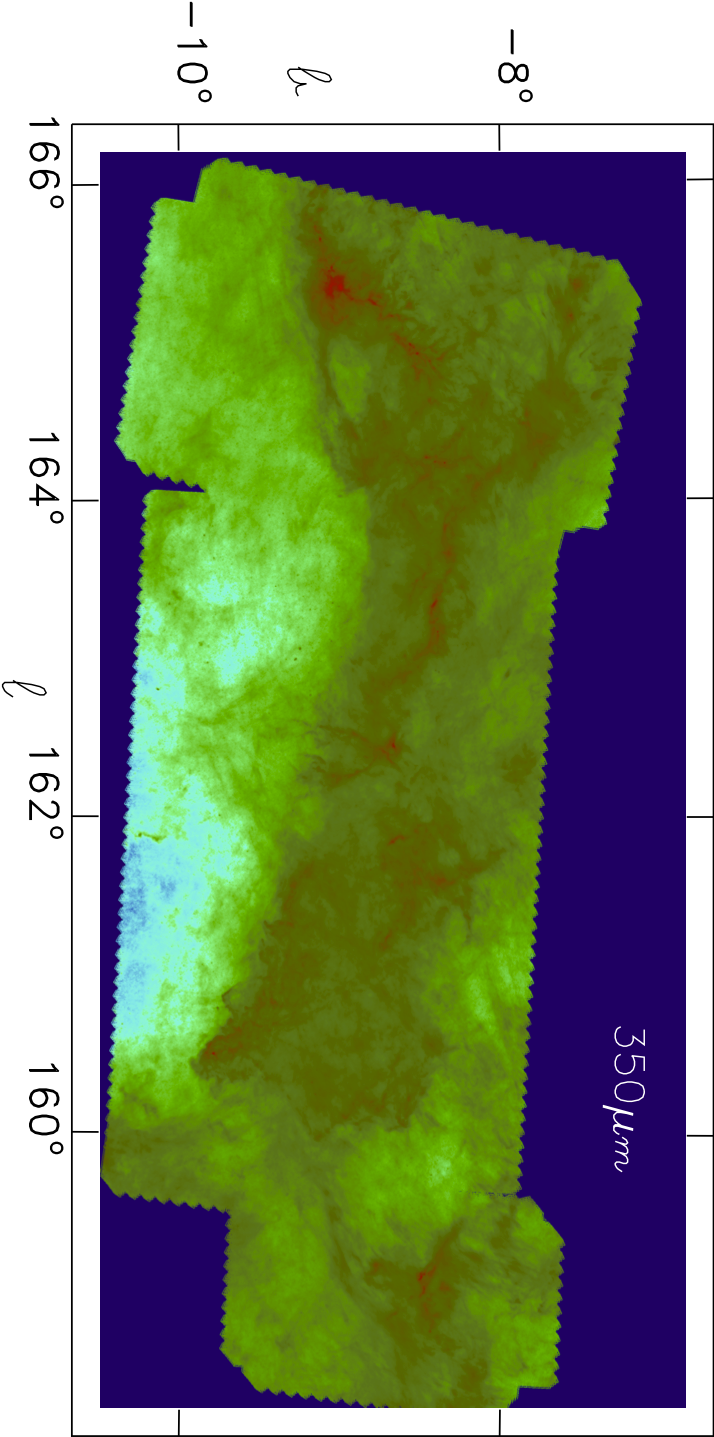}
\figcaption{\label{aur350}
False color image of the 350 \micron\ map derived with Scanamorphos from our Parallel-Mode observations.}
\end{figure}

\begin{figure}
\includegraphics[angle=90,totalheight=80mm]{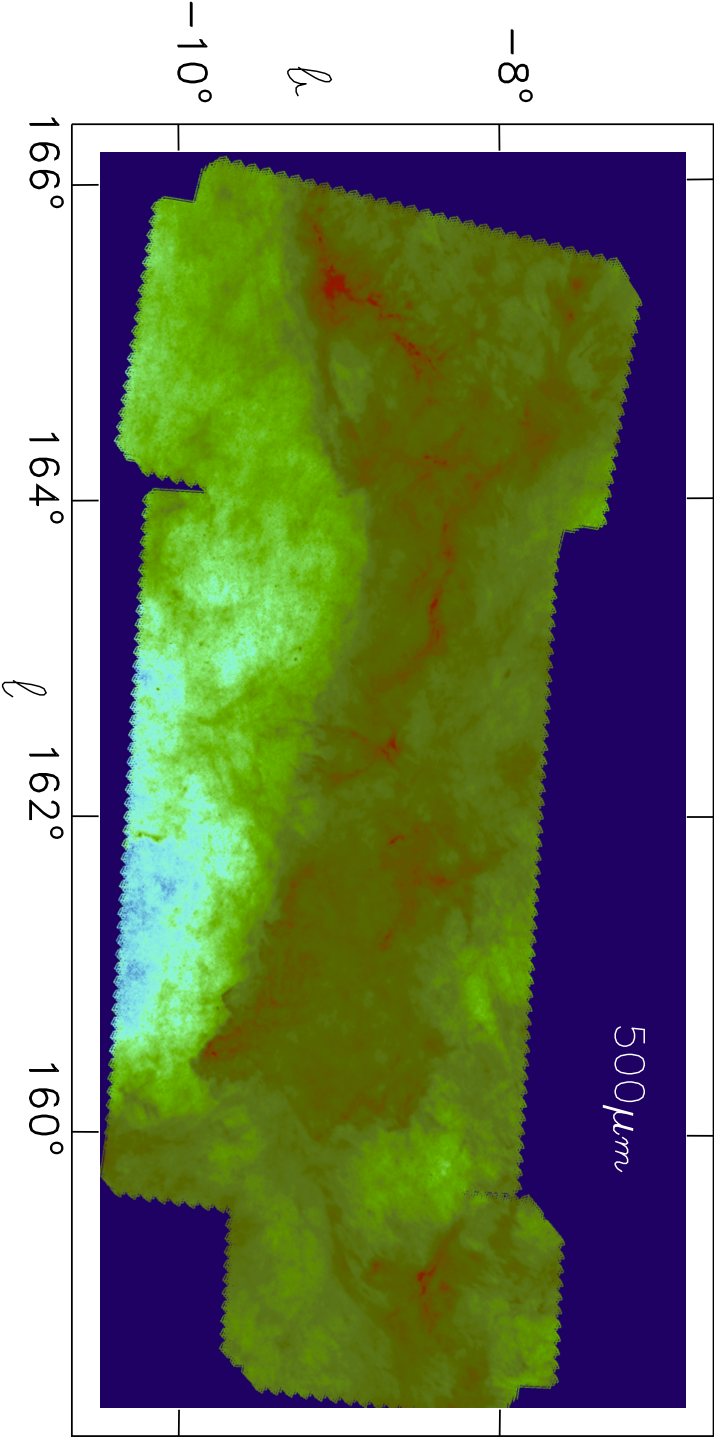}
\figcaption{\label{aur500}
False color image of the 500 \micron\ map derived with Scanamorphos from our Parallel-Mode observations.}
\end{figure}

The initial data reduction process is essentially the same as that used for several other star-forming clouds from the
Gould Belt Survey, e.g. \citep{sadavoy12,peretto12}.  The first step consists of reducing the {\it Herschel} data to
level 1 products using the Herschel Interactive Processing Environment (HIPE) version 8.1.0 \citep{ott10}.  Maps of the three
sub-regions listed in Table \ref{obsidtbl} were obtained 
using Scanamorphos version 16 \citep{scanam} 
using the two perpendicular scanmaps to remove correlated noise such as low frequency drifts.
The pixel scales for these maps were 3.2\as, 5\as, 6\as, 10\as, and 14\as\ respectively at 70 \micron, 160 \micron, 250 \micron,
350 \micron, and 500 \micron.
These individual maps are shown in Figures \ref{aur70}--\ref{aur500} (electronic edition only).
We then used two different source extractor routines. The first was the {\it getsources} package (version 1.120526)  \citep{getsources} that was developed to
search for sources over a range of spatial scales and extracts sources simultaneously over multiple bands that have
substantial differences in angular resolution.
The second source extractor was the {\it c2dphot} package developed as part of the
{\it Spitzer} Legacy {\it c2d} program \citep{harvey06,evans07} which was designed to work with point-like and small extended sources
up to roughly twice the beam size and was based on the earlier DOPHOT package \citep{dophot}.  
In this paper, we make use mostly of the results from the {\it c2dphot} processing (shown in Table \ref{fluxtbl}) since we are primarily addressing point-like
and very compact sources (\S \ref{compact}, \ref{indiv}) in addition to the very large scale structure (\S \ref{diff}).   Future publications will
use the results of the {\it getsources} processing to investigate the medium-scale emission.

\begin{figure}
\includegraphics[angle=90,totalheight=80mm]{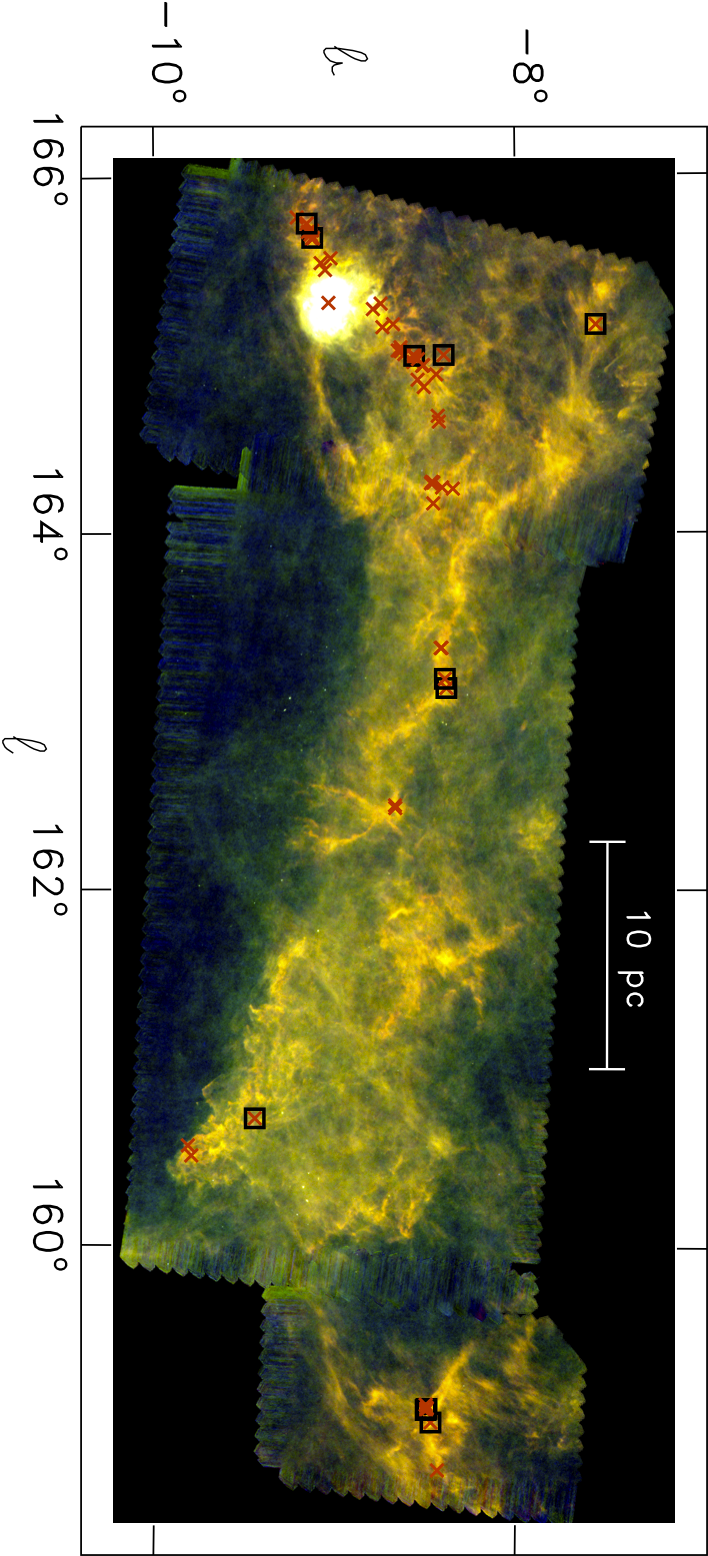}
\caption{\label{aurcalrgb}
False color image with 70 \micron\ (blue), 160 \micron\ (green), and 250 \micron\ (red) of the mapped area.  The locations of the
sources in Table \ref{fluxtbl} are marked with a red `X', and those that are not in the {\it Spitzer} YSO list of H. Broekhoven-Fiene et al. (2013, in preparation) are
also surrounded with a black square.}
\end{figure}

Figure \ref{aurcalrgb} shows a 3-color composite (70 \micron, 160 \micron, and 250 \micron) of the entire region mapped at 70 \micron, 160 \micron, 250 \micron, 350 \micron, and 500 \micron.
The two most obvious features of this map are: 1) the bright collection of sources and nebulosity at the left end of
the map (Southeast) where the LkH$\alpha$101 cluster is located, and 2) the long network of filamentary structure
that pervades much of the mapped area.  Such filamentary structure is now known to be typical in Galactic star-forming
regions from the work of the {\it Herschel} Gould Belt survey \citep{andre10} as well as the {\it Herschel} Galactic
Plane Survey, HIGAL \citep{molinari10} and has also been discussed earlier by \citet{myers09}.
Subsections of some of the mapped areas are discussed in more detail later in \S\ref{indiv}.

\begin{figure}
\includegraphics[angle=90,totalheight=150mm]{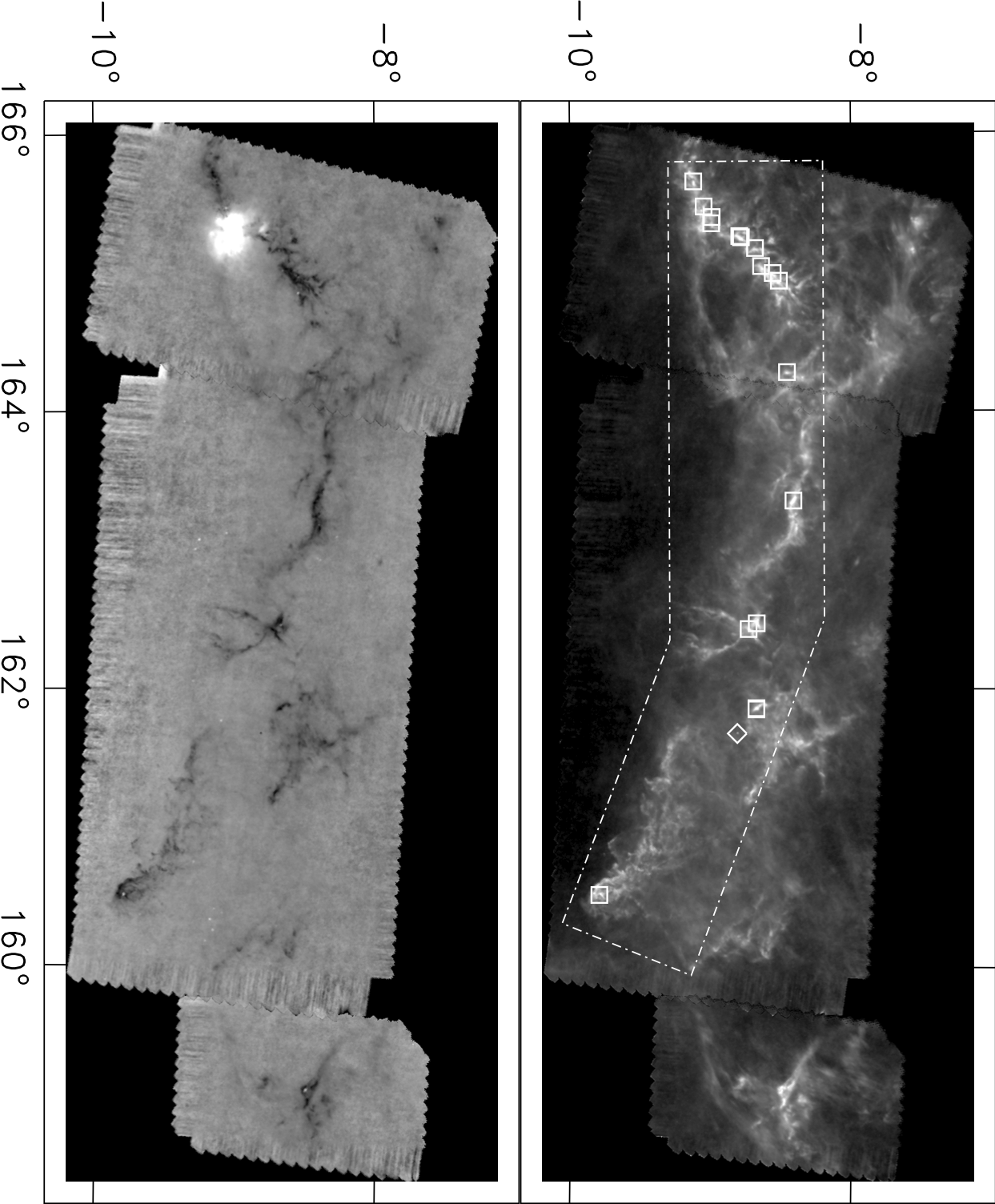}
\figcaption{\label{tempcold}
Upper image: column density image with positions of Bolocam 1.1 mm sources marked with squares (3C111 is indicated with the 
diamond). The area covered by our 1.1 mm Bolocam mapping
is outlined with the dash-dot line. The highest column areas (white) have
N$_{H2} \sim 5 \times 10^{22}$ cm$^{-2}$ while the lowest column areas represent values $\sim 1 \times 10^{21}$ cm$^{-2}$.  
Lower image: dust temperature image with maximum T$_d \sim$ 28K at LkH$\alpha$101 near the left
end of the image and minimum temperatures of order 10K in the darkest parts of the filaments.  The median derived dust temperature over
most of the area is $\sim$ 14.5 K.}
\end{figure}

In addition to the basic map-making and source extraction, we also present in Figure \ref{tempcold} results on dust temperature and optical depth
over the entire mapped area.   
We used a method similar to that described by \citet{kon10}; we first determined zero-point offsets following the procedure described by 
\citet{bernard10} and then convolved the shorter wavelength Herschel images to the resolution of the 500 \micron\ data.  We derived SED fits to the 
160 $\mu$m, 250 $\mu$m, 350 $\mu$m, and 500 $\mu$m data for each pixel position in the maps using a simplified model of dust emission, 
$F_\nu=\kappa_\nu \times B(\nu,T) \times$ column density.  
We assumed a dust opacity law of $\kappa_\nu=0.1(\nu/1000 GHz)^\beta cm^2/g)$ and fixed the dust emissivity index to $\beta$=2
with the standard mean molecular weight, $\mu$ = 2.33.
Because of the high S/N over most of the area of the flux maps that were used to derive these column-density and
temperature maps, the major uncertainty in the absolute values of T and $\tau$ are those due to the inherent assumptions
in using the equations above to represent the dust emission.  It is likely, though, that the maps provide an excellent
representation of relative temperatures and column densities with absolute uncertainties of order $\pm$ 15--20\% in
temperature and $\pm$ a factor of 2 or more in column density.
We discuss these maps more fully in \S \ref{diff} where we compare the column densities to those derived from extinction
measurements and analyze the distribution of star formation relative to the inferred gas densities (note, all column densities
discussed in this paper are measured as $N_{H2}$).

The digital versions of all these maps will be available soon after publication of this paper on the Herschel Science
Center's web site for user-provided data, 

http://herschel.esac.esa.int/UserProvidedDataProducts.shtml.

\subsection{CSO Observations}

We used the Bolocam imager\footnote{http://www.cso.caltech.edu/bolocam} at a wavelength of 1.1 mm to map much of the area covered in our {\it Herschel} observations
during the nights of 14--16 November 2011.
We utilized observing techniques similar to those used for the Bolocam Galactic Plane Survey (BGPS) as described by \citet{aguirre11} and
Ginsburg et al. (in prep.).
Alternating maps were made scanning roughly parallel and perpendicular to the Galactic plane at a scan speed of 120\as s$^{-1}$.
Multiple overlapping maps were obtained over roughly the eastern 2/3 of the {\it Herschel} mapped area; the total area observed was 6 deg$^2$;
the area covered is indicated in Figure \ref{tempcold}.
Due to non-uniform coverage and varying weather conditions the noise in the Bolocam maps is not constant, but is typically $\sim$ 0.07 Jy/beam.
This is substantially higher than the noise in maps of several other Gould Belt clouds presented by \citet{enoch07}, 0.01--0.03 Jy/beam, 
due to our significantly smaller observing time per pixel.  The primary flux calibrator was Uranus.
The map data were reduced using the software described by \citet{aguirre11} for the Bolocam Galactic Plane Survey (BGPS), utilizing
correlated sky-noise reduction with 3 PCA (Principal Component Analysis) components.  Following
that, sources were extracted as described by \citet{rosbgps} for the BGPS.
In addition to this large-scale mapping, we also observed a small area centered on one of the strong {\it Spitzer} sources
to the northwest of the scanned region, SSTGB04012455+4101490, for which we have no corresponding {\it Herschel} data.

\citet{aguirre11} have carefully investigated the inherent spatial filtering that occurs in removing correlated sky-noise in
ground-based observations at this wavelength.  For the case of subtraction of 3 PCA components roughly half the flux is
lost for structure larger than 300\as.  Indeed, the largest coherent area of 1.1 mm emission in our map is a 4\am\ wide
area centered on LkH$\alpha$101.  
Therefore, we present the results from this part of the study as positions and flux densities for the compact emission regions detected.
Table \ref{mmfluxtbl} lists these positions and the fluxes within several different apertures for the compact sources detected
at 1.1 mm with peak S/N $>$ 2.  Note, Source 2 is the bright galaxy 3C111 which was also our primary pointing calibrator and
secondary flux calibrator.  Table \ref{mmfluxtbl} also lists the {\it Herschel} sources from Table \ref{fluxtbl} that are located within 45\as\
of each 1.1 mm source position and likely associated with it.  We discuss these 1.1 mm sources below in \S \ref{1mmcompact}.

\section{Compact Sources}\label{compact}

\subsection{The 70 \micron\ Objects}

The goal of our investigation of the compact sources in the AMC is to complete the search for
pre-main-sequence and proto-stars that began with the {\it Spitzer} Gould Belt program (H. Broekhoven-Fiene et al. 2013, in preparation) and, in particular,
to search for the most dust-enshrouded objects that might have been missed by that program because they emit most
of their luminosity in the far-IR.  {\it Herschel} at 70 \micron\ provides the highest
resolution imaging in the far-IR of any current or planned facility, and conveniently the 70 \micron\ resolution ($\lambda/D \sim$ 4\as) is also nearly identical
to that of {\it Spitzer} at 24 \micron.  Although the resolution of Parallel-Mode observations is not quite
as high as {\it Herschel's} diffraction limit because of image blur from the fast scan speed in Parallel-Mode,
the resolution achieved is not much below that limit.  Therefore, an additional goal of this investigation is
to use this resolution to measure fluxes in the far-infrared more reliably than {\it Spitzer} in confused regions.
With a complete and reliable census of all the stages of star-formation in the AMC, we will be able to
make the most informative comparison of it with the OMC.

The source extractor {\it c2dphot} operates in two modes.  In the first mode, it searches through the image
at sequentially lower flux levels for local maxima, characterizes them as point-like or extended (ellipsoidal), and
subtracts them from the image.  In the second mode the code is given a list of fixed positions at which it
fits the point-source-function (PSF) to whatever flux above the background exists at that position.
This mode is useful for determining upper limits and for testing for faint objects in the wings of bright ones.
In both modes an aperture flux is calculated as well as the PSF- or ellipsoidal shape-derived flux.
To find the most complete set of possible objects to correlate with objects at other wavelengths, we first
processed both the 70 \micron\ and 160 \micron\ images in the most general {\it c2dphot} mode, allowing the code to fit flux, position, and shape
down to the lowest flux levels present in the image, i.e., essentially the noise level.  This process produced a list of
$\sim$6500 sources at 70 \micron\ and $\sim$500 sources at 160 \micron, of which probably over half are noise at both
wavelengths.
After comparing a number of individual cases while trying to correlate the 160 \micron\ objects with those found
at 70 \micron, we identified two complicating issues.  First, the obvious issue of the larger PSF at the longer
wavelength meant that sometimes more than one 70 \micron\ source would be within the 160 \micron\ PSF.   Second,
because cooler, more extended dust is naturally detected at the longer wavelength, in some cases the 160 \micron\ source
equivalent to a nearly point-like 70 \micron\ source would be extended and have an asymmetric shape, making an automated
detection and association difficult.  For these reasons we decided to determine the 160 \micron\ fluxes (or limits in most
cases) for the 70 \micron\ detections by running {\it c2dphot} in its second, fixed-position mode at 160 \micron, using the
70 \micron\ detection list for the input positions.  In this case it can also be useful to compare the PSF-fit fluxes with
those determined from aperture photometry as a secondary indication of larger source extent or confusion.
For the six coldest sources discussed later in Figure \ref{srcsed} with T$_{bol} < $ 40K we have also extracted flux densities at 250 \micron\ 
in the same way as the 160 \micron\ fluxes and show the PSF-fit fluxes.  Since the 250 \micron\ PSF has a full-width-half-maximum
of roughly 18\as, we have not measured the fluxes of the bulk of our objects beyond 160 \micron; the issues with assigning fluxes
to individual sources at 160 \micron\ are incrementally more problematic at the SPIRE wavelengths.  A future study making use
of the {\it getsources} processing is likely to produce the most reliable long wavelength SEDs for most of the sources.

\begin{figure}
\includegraphics[angle=180,totalheight=120mm]{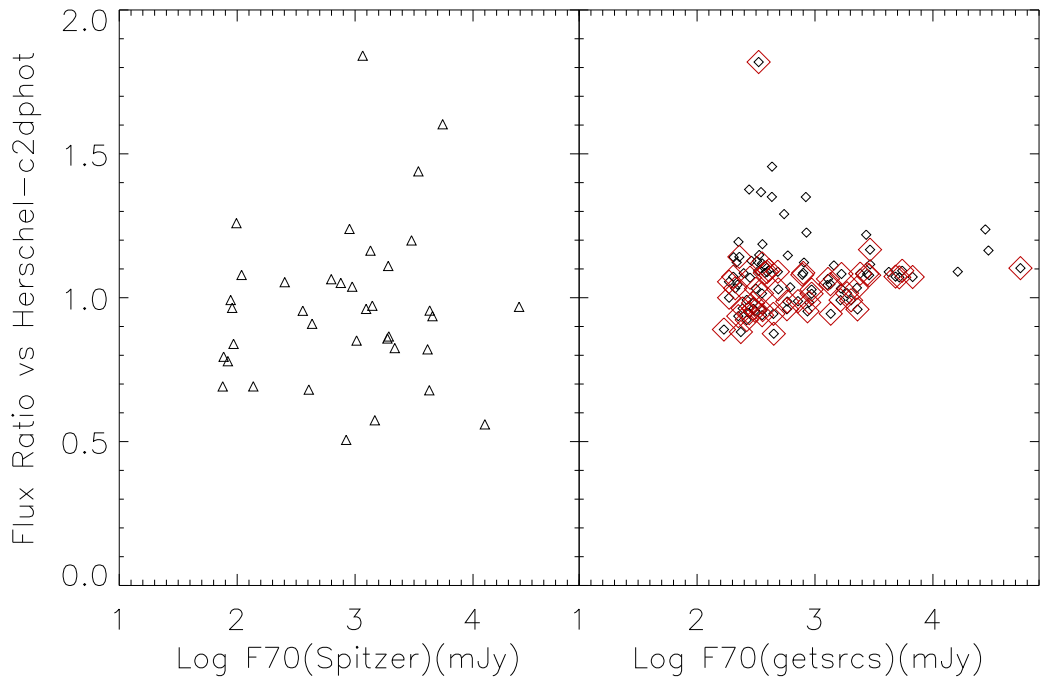}
\figcaption{\label{fluxcompare}
Comparison of 70 \micron\ fluxes (mJy) derived in this study. Left panel: ratio of the fluxes from the {\it Spitzer} Gould
Belt survey to the {\it Herschel c2dphot} fluxes from this study for the
YSOs in common. Right panel: ratio of the fluxes using the {\it getsources} algorithm with our {\it Herschel} maps to those
from our {\it c2dphot} processing for the
high S/N objects found not to be extended on the basis of the quality of fit or derived source size. Those marked with red diamonds also
included a criterion on ratio of total flux to PSF-fit flux being less than 1.5. }
\end{figure}

The nominal absolute calibration uncertainty in PACS photometry is now believed to be $\pm$3 \% at 70\micron\ and $\pm$ 5\%
at 160\micron\footnote{http://herschel.esac.esa.int/twiki/bin/view/Public/PacsCalibrationWeb?template=viewprint}.
These values only apply to well-sampled, color-corrected point sources.   
The SPIRE photometry is believed to be calibrated to $\pm$ 5\% under equally ideal circumstances\footnote{http://herschel.esac.esa.int/hcss-doc-9.0/}.
For the purposes of this study,
we assume the more conservative value of $\pm$ 15\% as used, for example, by \citet{kon10} from the original instrument papers by
\citet{pacs10} and \citet{spire}.
Given that we have flux measurements for a number of sources at 70 \micron\ also from the {\it Spitzer} Gould Belt program as
well as flux determinations from our {\it Herschel} data set using the completely different {\it getsources} algorithm,
we have an opportunity to check our flux measurements for systematic effects and problems like non-linearity.
Figure \ref{fluxcompare} shows plots of these two comparisons.   For the {\it Spitzer} comparison we used the list of
reliable YSOs described by H. Broekhoven-Fiene et al. (2013, in preparation) and excluded several objects in very confused regions.  The {\it Spitzer} photometry was
produced by the version of {\it c2dphot} described in detail by \citet{evans07} for the final delivery of {\it c2d} data to
the Spitzer Science Center.   For the best {\it getsources}
comparison (red diamonds in Figure \ref{fluxcompare}) we used only sources found to be isolated, well-fit by a point source at 70 \micron, 
with a product of major axis and minor
axis less than 150\as$^2$, a total flux less than
$1.5 \times$ the point-source flux, and a $S/N > 10$.  
The mean ratio of {\it Herschel} to {\it Spitzer fluxes} is 1.03$\pm$0.3, and the mean ratio of {\it c2dphot} fluxes to
{\it getsources} fluxes is  0.96$\pm$0.13 for the sources marked with red diamonds.  The scatter between the {\it Herschel} and {\it Spitzer} results appears
to be independent of flux level, while that between the two methods used on the {\it Herschel} data is consistent with
what one would expect as a function of the signal-to-noise level.
This excellent agreement, particularly for the two methods used on the {\it Herschel} data,
between flux determinations over a very wide range of brightness suggests that both {\it c2dphot} and {\it getsources}
provide highly reliable extractions and flux determinations, certainly for compact objects.

To identify the objects in the AMC that are most likely to represent young sources with a
stellar or pre-main-sequence core, we culled our starting list of $\sim$ 6500 sources in several ways using shorter wavelength
data.  Much of our observed area has been covered at 24 \micron\ in the {\it Spitzer} Gould Belt survey of H. Broekhoven-Fiene et al. (2013, in preparation); in areas
not observed with {\it Spitzer} the recently released WISE all sky survey \citep{wise} provides relatively deep mid-infrared
photometry over a similar range of wavelengths.
To start our search for reliable young objects we  included only those sources with
S/N $> 7.5$ and $F_\nu$ at 70 \micron $>$ 85 mJy with at least a $2.5 \sigma$ detection at one of the closest neighboring wavelengths, i.e.
22/24 \micron\ Spitzer-MIPS/WISE, or our 160 \micron\ {\it Herschel} photometry.  These criteria reduced the list of possible
young objects to 513 sources.  \citet{koenig12} have estimated a contamination rate for extragalactic sources in star-forming regions
at the sensitivity and wavelength of the WISE survey of $\sim$ 10 objects per square degree.  This high background level suggests that a significant
fraction of our 513 candidate sources is extragalactic.  

\begin{figure}
\includegraphics[angle=0,totalheight=180mm]{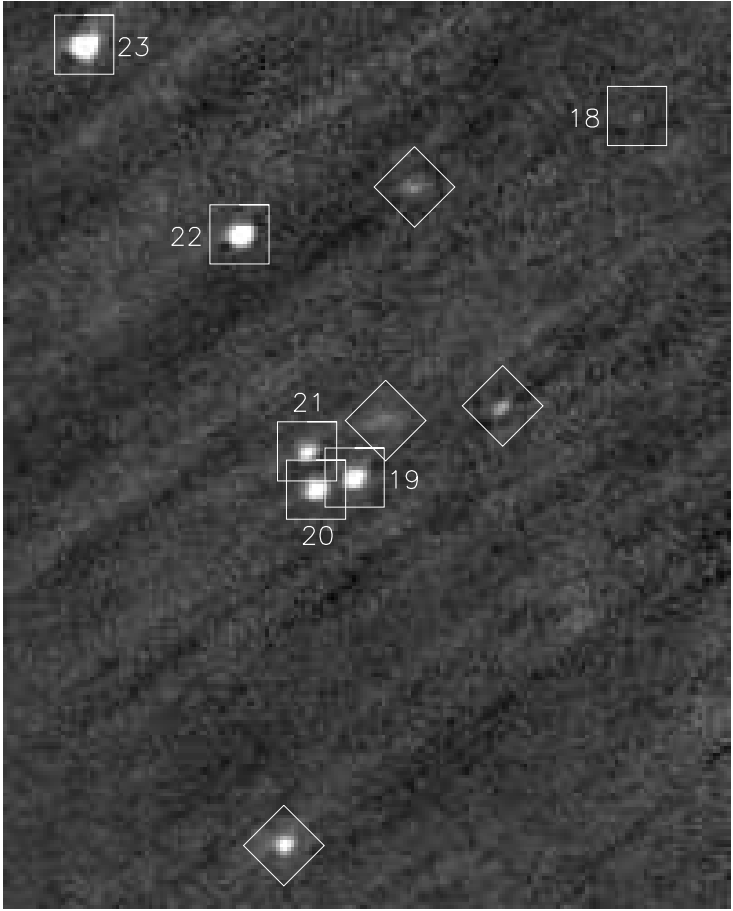}
\figcaption{\label{galexamp}
Field (10\am\ $\times$ 13\am) centered at J2000 R.A. 04$^h$ 28$^m$ 33$^s$, decl. +36\degree\ 25\am\ 20\as\ with 6 sources marked on the 70 \micron\ image
from Table \ref{fluxtbl} with squares and their source number, and 4 objects with diamonds identified
as extragalactic using the 2MASS ``gal\_contam'' flag.  Orientation is North-up, East-left.}
\end{figure}

To eliminate as many extragalactic sources from the candidate list as possible, we used a multi-pronged
approach that relied on: (1) examination of individual images for likely galaxies at 70 \micron\ as well 
as 24 \micron\ ({\it Spitzer}), 2MASS, and the red
(DSS) Digitized Sky Survey images, (2) {\it Spitzer/WISE} color-color diagrams using the criteria developed by \citet{koenig12}, (3) the
2MASS ``gal\_contam'' flag which signifies a likely extended extragalactic object (gal\_contam = 1), and (4) objects of low S/N in
the images or those that had no clear point-like core at 70 \micron.  Because of the wide range in colors, brightnesses, and angular sizes of the
extragalactic objects, all of these criteria contributed substantially to the elimination process.   After this triage we were left with 209 possible young
candidates.  This sample was clearly still ``polluted'' with extragalactic sources and evolved stars as we found by
searching the SIMBAD data base and examining several dozen sources in DSS images.  The most likely YSO candidates found by
{\it Spitzer} (H. Broekhoven-Fiene et al. 2013, in preparation), however, are located exclusively in the areas of high column density illustrated in Figure \ref{tempcold},
as are the 1.1 mm sources found in our Bolocam survey.   Most of our 209 70 \micron\ candidate young objects, though, are distributed
quite uniformly over the area, as were the previously culled extragalactic objects and candidates.  
Therefore, at the risk of missing a likely very small number of
young objects outside the high-column-density areas, we decided to apply one more criterion to our search list, i.e.,  to require
the column density, as measured by {\it Herschel} at the 35\as\ resolution of the column density map,
to be above 5 $\times 10^{21}$ cm$^{-2}$ (N$_{H2}$) at the source position.  This threshold
reduced the young candidate list to 60 objects.  To be confident that this criterion did
not eliminate any cold, dense young sources, we examined the DSS and 2MASS images of the few objects in lower column
density regions that had $F_{160}/F_{70} > 1.5$ and $F_{70} > 150$ mJy, and all appeared extragalactic, i.e. not point sources.  Therefore, the only young objects
that we may have missed would be relatively faint and blue.
Indeed, within the sample of {\it Spitzer} YSOs (H. Broekhoven-Fiene et al. 2013, in preparation) only one out of the $\sim$60 objects in the Class I--II range
lies outside the N$_{H2} \ge 5 \times 10^{21}$ cm$^{-2}$  area of our column density map.
The combination of all of the above criteria for eliminating extragalactic objects was important in reaching our final sample; for example,
if we had simply applied the column density criterion alone, we would have extracted a sample of $\sim$ 120 objects, half of which obviously
would have been background extragalactic sources behind higher column density local material.
As an example of the contamination issue from extragalactic objects,
Figure \ref{galexamp} shows a small field in the filament north of LkH$\alpha$101 with 6 YSOs
identified by H. Broekhoven-Fiene et al. (2013, in preparation) (squares) and 4 objects identified as extragalactic from the 2MASS ``gal\_contam'' flag (diamonds) that
are also extended when examined carefully at 70 \micron.

Table \ref{fluxtbl} lists the positions and 70 \micron/160 \micron\ {\it c2dphot} flux determinations of the final list of 60 objects found at 70 \micron\ that
appear to be reliable young members of the AMC.  The uncertainties listed are the statistical uncertainties of the measurements
only; the absolute calibration uncertainties of $\pm$15\% have already been mentioned.
For those sources that are identified also in the {\it Spitzer} Gould Belt data set, the 24 \micron\ Spitzer flux
is given, and if that is not available, the 22 \micron\ WISE \citep{wise} flux is listed if available.  The objects that
are identified as YSO candidates by H. Broekhoven-Fiene et al. (2013, in preparation) are indicated, and where possible we list the
object type shown by SIMBAD.  Note, we think it unlikely that the 4 objects listed as ``PN?'' are truly evolved objects based
on both their photometry and location in the AMC cloud.  
Table \ref{fluxtbl} also lists the spectral slope $\alpha$ ($\alpha = {dlog(\lambda F(\lambda)) \over dlog(\lambda)}$) 
determined from all existing members of the set of 3.6 \micron, 24 \micron, 70 \micron,
and 160 \micron\ photometry\footnote{Note: the original definition of ``$\alpha$'' used photometry only out to 24 \micron, 
so our spectral slopes are not directly comparable to earlier measurements in many cases.}; also shown is 
the corresponding YSO class based on the nomenclature of
\citet{lada87} as extended by \citet{greene94} as well as the total luminosity  and the bolometric
temperature as defined by \citet{myers93} determined over the same 3.6 \micron--160 \micron\ wavelength range.
We have extended the classification to ``Class 0'' to signify the most dust-enshrouded objects as suggested by \citet{andre93}.  Since we do not have
high angular resolution photometry for most of the objects at $\lambda \sim$ 1~mm, we have used the bolometric temperature to identify these
objects within the nominal Class I category as defined by spectral slope.  We used the criteria that objects with T$_{bol} \le$ 50 K are 
likely candidates for Class 0 objects, and those with 50~K~$< $T$_{bol} \le$~70~K
we have marked as Class I/0, since some of them are likely to be identified as Class 0 when the requisite submillimeter photometry exists and reliable 
submillimeter to bolometric luminosity ratios can be derived.
The class determinations are generally similar to those found for the YSO candidates
of H. Broekhoven-Fiene et al. (2013, in preparation), though the addition of our more reliable 70 \micron\ and 160 \micron\ photometry has made changes for a few.

H. Broekhoven-Fiene et al. (2013, in preparation) report a total of 164 YSO candidates based on {\it Spitzer} and WISE photometry at $\lambda \le$ 24\micron\ within the
area covered by our {\it Herschel} survey.  Clearly
a significant number of these are not detected in our study.  We have examined this list of non-detections and find that most of them
are simply too faint and blue to be likely to be detected in the far-infrared at our sensitivity level.  Several redder and brighter
{\it Spitzer} YSO candidates lie within the confused region around LkH$\alpha$101.

There are 11 objects in Table \ref{fluxtbl} that are not in the {\it Spitzer} YSO candidate lists of H. Broekhoven-Fiene et al. (2013, in preparation).
Two of these (Sources 4 and 42) are completely undetected at 22/24 \micron\ by WISE/{\it Spitzer}. We {\it were} able, however, to derive
rough upper limits from the existing MIPS 24 \micron\ images of 2 mJy for source 4 and 3 mJy for Source 42.
Clearly both these objects exhibit relatively cold spectral energy distributions (SEDs).  
Most of the remaining 9 objects not found as YSO candidates
by H. Broekhoven-Fiene et al. (2013, in preparation) were not selected by them simply because at least one IRAC or WISE band was missing from the detection list making
classification as a YSO impossible using the {\it Spitzer c2d/GB} criteria.   One of the 11 objects not identified previously as a YSO is
Source 49 which has a very blue SED in the IRAC bands, but very strong far-infrared emission.  
It is one of only 6 objects with
$T_{bol} <$ 40K (see also Figure \ref{rgbysotot} and \S \ref{indiv}).



\subsection{The Bolocam Sources}\label{1mmcompact}

Figure \ref{tempcold} shows the location of the 18 1.1 mm sources from Table \ref{mmfluxtbl} that are within our {\it Herschel} area
relative to the column density
derived from our {\it Herschel} mapping.  With the exception of the galaxy 3C111 all of the 1.1 mm peaks fall on
filaments of high column density.   Excluding 3C111 and LkH$\alpha$101, $3/4$ of the 1.1 mm sources have {\it Herschel} 70 \micron\ objects associated
with them, and these are generally from the two ``earliest'' YSO classes,  0 and I (Note: Bolocam source 1 is associated with 70 \micron\ source 9,
which is a very bright emission-line star that is a likely FU Orionis object \citep{sand98}).  There are, however, four 1.1 mm emission
sources with no associated {\it Herschel} source from Table \ref{fluxtbl}.  Source 10 is associated with a diffuse ``blob''
of emission near LkH$\alpha$101 in our 70 \micron\ map, but Sources 3, 4, and 5 have no clear 70 \micron\ counterpart.  They are
also three of the 1.1 mm sources with the lowest S/N ratio, but as just noted, they are clearly located on high-column-density filaments.

\begin{figure}
\includegraphics[angle=0,totalheight=180mm]{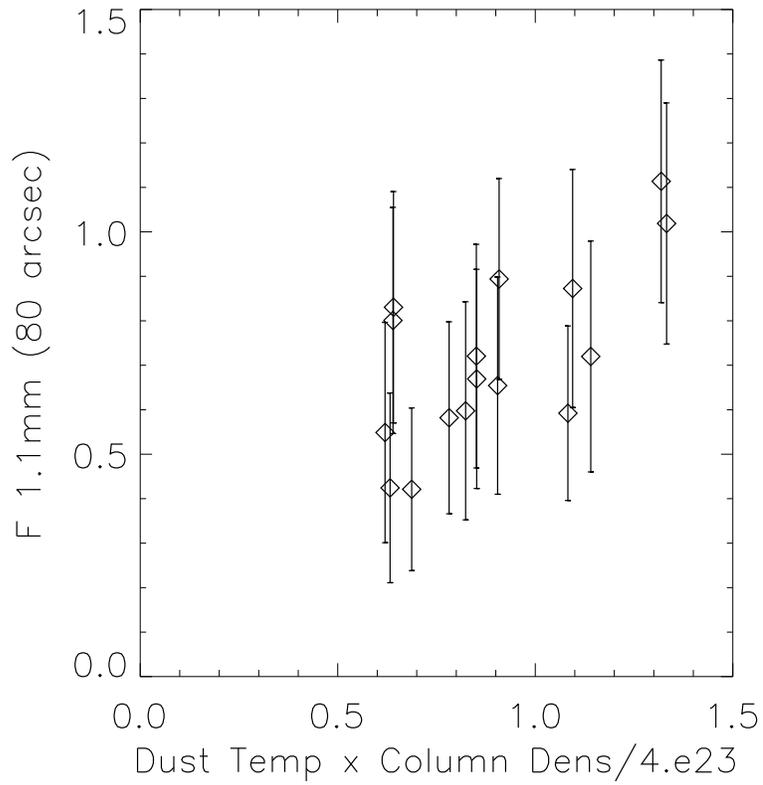}
\figcaption{\label{csofluxcol}
Plot of 1.1 mm flux density in an 80\as\ aperture from Table \ref{mmfluxtbl} versus product of {\it Herschel} dust temperature and column density averaged over the
same size aperture.  LkH$\alpha$101 and 3C111 are not included.  The correlation coefficient for this distribution is 0.67.  The formal uncertainties
in the product of dust temperature and column density are insignificant in comparison to the uncertainties due to the particular model used.}
\end{figure}

If the bulk of the 1.1 mm emission from the sources in Table \ref{mmfluxtbl} arises from very cool, dense, dust, then we might
expect a correlation between the 1.1 mm flux density and the total {\it Herschel}-derived dust column density at that point, or more likely, the product
of the {\it Herschel}-derived column density and temperature.  This correlation would likely exist whether the dust is heated internally by a compact pre-main-sequence
object or externally by the interstellar radiation field.  To test this idea we have plotted in Figure \ref{csofluxcol} the 1.1 mm
flux in an 80\as\ aperture versus the product of the {\it Herschel}-derived column density and temperature at the position of the 
1.1 mm source, averaged over 80\as.   We have not included the galaxy 3C111 nor the hot, luminous source associated with LkH$\alpha$101
whose core dust temperature is unlikely to be well-sampled with the {\it Herschel} beams and whose flux levels are saturated
at several {\it Herschel} wavelengths.  Figure \ref{csofluxcol} shows a
rather good correlation between 1.1 mm flux and the product of the {\it Herschel}-derived dust column density and temperature for all the other 1.1 mm
sources.  Thus, the three 1.1 mm sources without associated compact 70 \micron\ sources  probably deserve future investigation as possible
pre-stellar cores\footnote{Bolocam sources 3 and 4 are positionally associated with the bright F star, HD 27214, but the Hipparcos
parallax for this star of 11.51 milli-arcsec  implies that HD 27214 is much closer than the AMC.}.

\section{Individual Sources}\label{indiv}

\begin{figure}
\includegraphics[angle=0,totalheight=180mm]{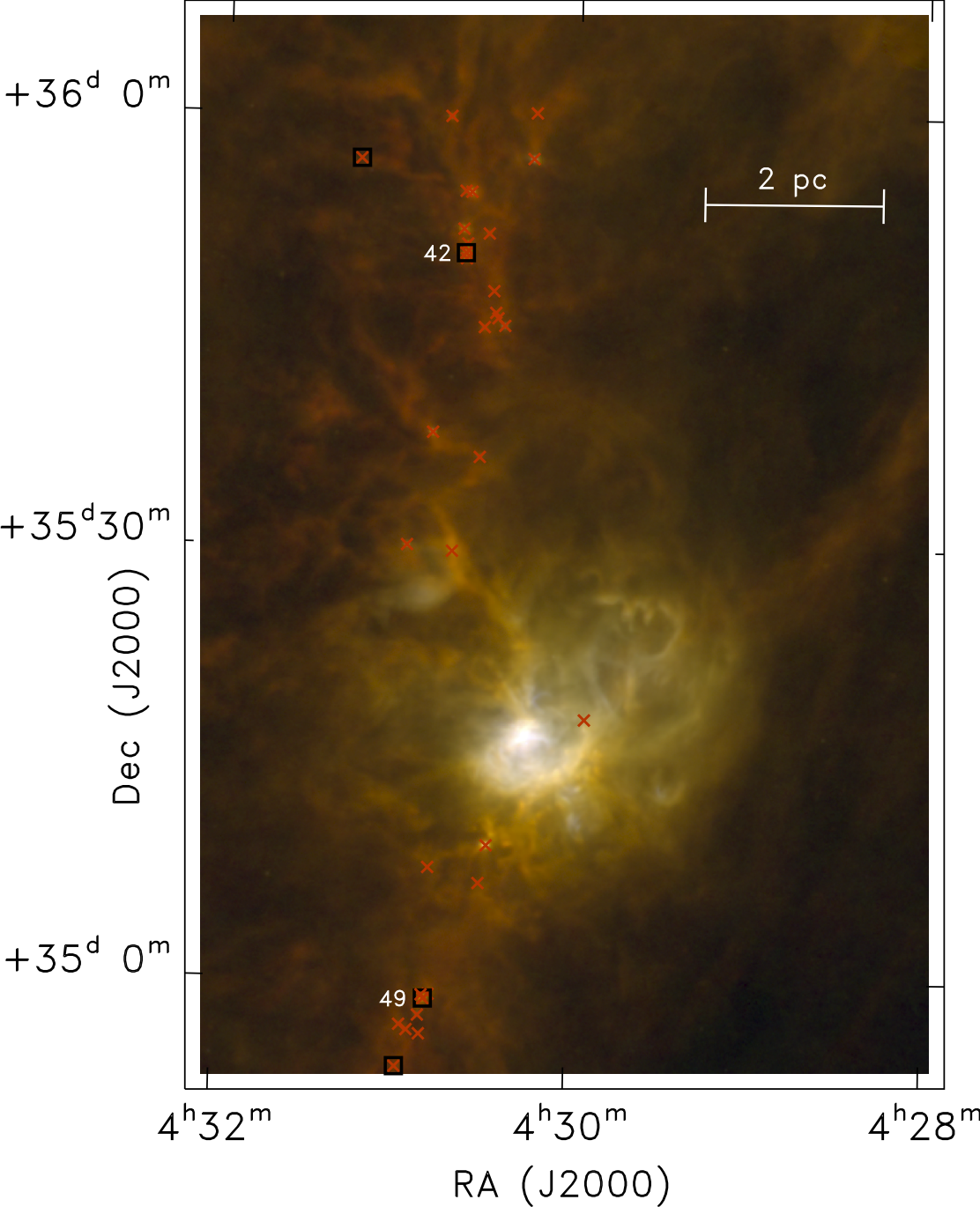}
\figcaption{\label{lkhargb}
False color image with 70 \micron\ (blue), 160 \micron\ (green), and 250 \micron\ (red) of the LkH$\alpha$101 area and star-forming
filament to its north.  
As for Figure \ref{aurcalrgb}, the locations of the
sources in Table \ref{fluxtbl} are marked with a red `X', and those that are not in the {\it Spitzer} YSO list of H. Broekhoven-Fiene et al. (2013, in preparation) are
also surrounded with a black square.
LkH$\alpha$101 is the brightest object in the image.  
The second black square down from the northern boundary of the image is the location of Source 42, shown in more detail in Figure \ref{rgbysotot}.  
The black square at the bottom of the image marks Source 49, also shown in more detail in Figure \ref{rgbysotot}.
}
\end{figure}

There are several regions within our {\it Herschel} maps that are notable for either the number of young objects or 
strong emission at the longer wavelengths.  H. Broekhoven-Fiene et al. (2013, in preparation) note the large number of {\it Spitzer} YSO candidates near LkH$\alpha$101
and in the filament extending north and west of it by roughly 1 degree.  Figure \ref{lkhargb}  shows a 3-color composite image
of this area with the positions of the {\it Herschel} sources from Table \ref{fluxtbl} marked.  Roughly 60\% of the likely young
far-infrared sources listed in Table \ref{fluxtbl} are within a  1.2\degree\ $\times$ 0.5\degree\ area (9.4 pc $\times$ 3.9 pc) centered on the high-column-density
filament shown in this area in Figure \ref{tempcold}.  The YSO population in the roughly 4\am\ $\times$ 4\am\ core of this region 
(0.5 pc square) centered on LkH$\alpha$101
has been summarized by \citet{herbig04} and \citet{sfhb08}. More recently \citet{guter09} discussed {\it Spitzer} observations
of this cluster in comparison with a number of other
young clusters and \citet{wolk10} have added x-ray data to further define the cluster properties. 
Our {\it Herschel} observations are complementary to these studies in that the central few arcminutes of
our images are dominated by the diffuse dust emission, presumably heated by the central bright star as well as the dense
population of lower luminosity stars surrounding it.  Beyond a radius of
$\sim$ 3\am\ from LkH$\alpha$101, though, we are sensitive to compact thermal emission from individual members of the extended YSO population.

\begin{figure}
\includegraphics[angle=0,totalheight=130mm]{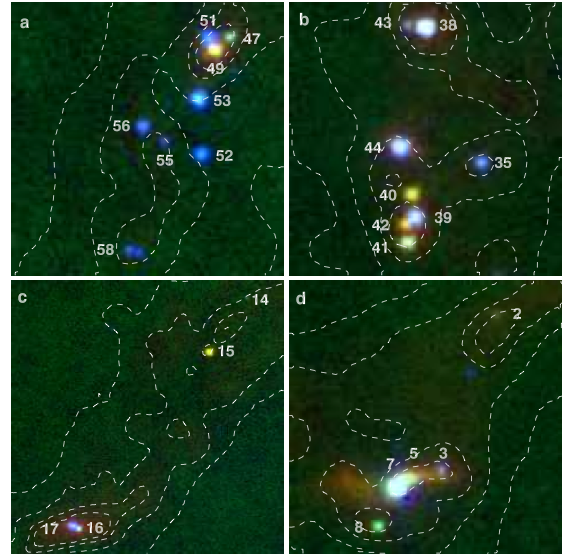}
\figcaption{\label{rgbysotot}
False color images of the four areas discussed in \S \ref{indiv} with 24 \micron\ (blue), 70 \micron\ (green), and 160 \micron\ (red).  The sources
are labeled with their reference numbers from Table \ref{fluxtbl} (for clarity, sources 4 and 6 are not marked in Panel d).
Column density contours from the map shown in Figure \ref{tempcold} are overlaid at values of 0.5, 1.0, 2.0, and
3.0 $\times\ 10^{22}$ cm$^{-2}$ ($N_{H2}$).  Panels a--d respectively have angular sizes of 6.4\am, 5.9\am, 12.8\am, and 6.9\am. }
\end{figure}

\begin{figure}
\includegraphics[angle=0,totalheight=130mm]{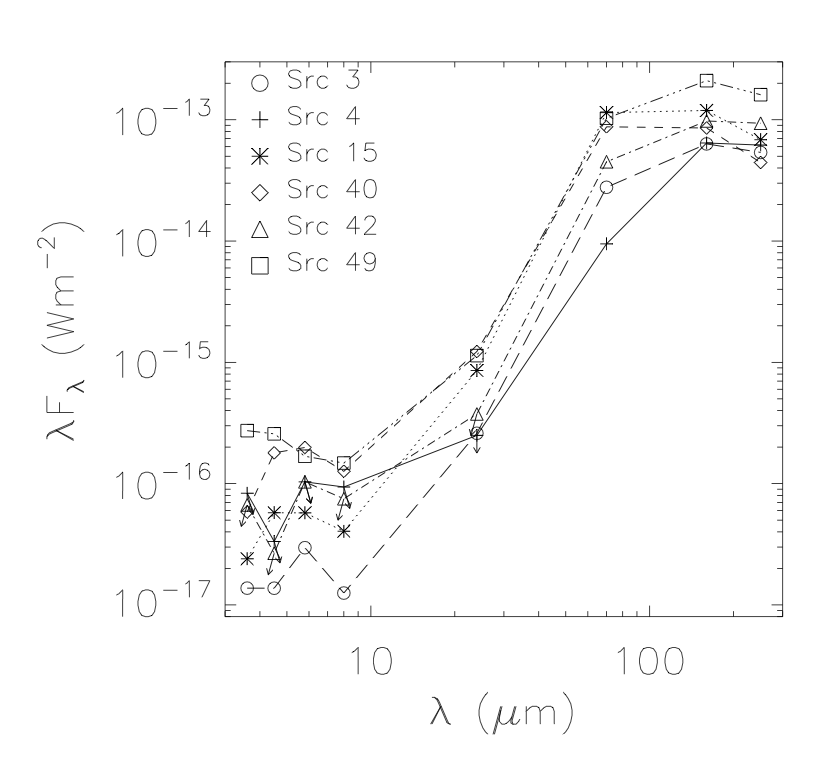}
\figcaption{\label{srcsed}
Spectral energy distributions of the six coldest sources discussed in \S\ref{indiv}. Note, sources 4 and 42 have only flux upper
limits shortward of 70 \micron.}
\end{figure}

At the southern end of the filament containing LkH$\alpha$101 is a group of sources that includes Source 49, mentioned earlier
as one of the objects with a very low T$_{bol}$ in spite of being detected easily by {\it Spitzer's} IRAC in all four bands
with relatively blue colors.  It also is extended in the IRAC images on a scale of $\sim$2--3\as\ in the north-south direction. 
These characteristics are consistent with it being a disk viewed edge-on, where the IRAC emission arises from scattered light
above and below the plane of the disk along low extinction lines of sight to the central star; other interpretations are, of course, also possible.
An expanded view of this source along with six other objects is shown in Figure \ref{rgbysotot}a.  The derived column density in this area peaks
at $4 \times 10^{22}$ cm$^{-2}$ essentially at the position of Source 49, and this area is associated with Bolocam Source 18.
The SEDs of Source 49 and several other of the coldest sources discussed below are shown in Figure \ref{srcsed}. As mentioned
earlier, for these six
examples of the coldest source SEDs we have extracted approximate PSF-fit fluxes at 250\micron\ which show that the SEDs of all of these objects
peak shortward of 250\micron.  Therefore, we have not found any compact objects with extremely cold SED's, although this may be related
to our requirement for detectable 70\micron\ emission.

Further north along the filament is a tight collection of young objects that includes one of our two 70 \micron\ sources
that was undetected by both {\it Spitzer} and WISE in the mid-infrared, Source 42 in Table \ref{fluxtbl} with spectral
slope $\alpha \sim 3$ and also associated
with Bolocam 1.1 mm Source 14.  A closeup of this region is shown in Figure \ref{rgbysotot}b.  The image also includes  70 \micron\ Sources 
35, 38, 39, 40, 41, 43, and 44.

More than a degree northwest of the filament containing LkH$\alpha$101 is another region of high column density
containing four objects from Table \ref{fluxtbl}, Sources 14 -- 17.  Source 16 is a pointlike Class I/0 object
located $\sim$ 16\as\ west of Source 17 (Class II), the latter of which is elongated in the direction of Source 16.  Both
are also associated with Bolocam Source 7, the third brightest 1.1 mm emission region.  These two objects are shown in
the lower left of Figure \ref{rgbysotot}c.  In the upper right are Source 15, another very cold object with spectral 
slope $\alpha \sim 3$, and Source 14, a faint Class I object.

Finally, in Figure \ref{rgbysotot}d we show a tight collection of 7 Class 0 and I sources at the northwest end of our
mapped area.  These sources include an extended object at {\it Herschel} wavelengths
that has been extracted as five separate condensations at 70 \micron\ (Sources 3--7), three
of which are well isolated {\it Spitzer} YSOs from H. Broekhoven-Fiene et al. (2013, in preparation).  At 160 \micron, however, the five sources are blended into a
single elongated structure that peaks on the position of the brightest source at 24 \micron, 70 \micron, and 160 \micron.  Sources 2 and 8 from Table \ref{fluxtbl}
are also included in this figure.

\section{Cloud Mass and Structure}\label{diff}

We have performed a preliminary analysis of the diffuse dust emission as described earlier by fitting the 160 \micron--500 \micron\
emission with a simple SED that characterizes the dust with a temperature and column density.  These results shown in Figure
\ref{tempcold} reveal a network of narrow filaments characterized by column densities of up to a few $\times\ 10^{22}$ cm$^{-2}$ ($N_{H2}$)
and temperatures that drop to $\sim$ 10 K from the typical value of order 14--15K in the low-density parts of our maps.
Many of these filaments are associated with Lynds dark clouds as indicated in Figure 1 of \citet{lada09}.
In an initial effort to quantify the differences and similarities between the AMC and other star-forming regions we have calculated
two quantities discussed by other authors for similar clouds. These quantities are the cumulative mass fraction as a function of extinction as
already mentioned by \citet{lada09} for the AMC and the probability density function for the column density as discussed
by \citet{schneider12} and others.  

\begin{figure}
\includegraphics[angle=0,totalheight=100mm]{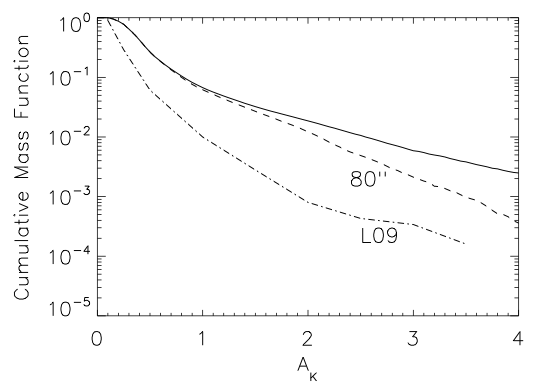}
\figcaption{\label{cummsf}
Normalized cumulative mass fraction for the area observed versus K magnitude extinction using the same assumptions as those of \citet{lada09} (solid line).
The dash-dot line shows the function derived by \citet{lada09} with the NICER technique as read from their Figure 4, which is reported as
smoothed with an 80\as\ HPW Gaussian.  The dashed line shows the result of smoothing our {\it Herschel}-derived column density map with an 80\as\
HPW Gaussian.}
\end{figure}

Figure \ref{cummsf} shows the cumulative mass fraction versus K magnitude extinction A$_K$ using the same conversion
factors as \citet{lada09}, 2 $\times$ N$_{H2}$/A$_K$ = 1.67 $\times$ 10$^{22}$ cm$^{-2}$ mag$^{-1}$.  
The figure also shows the Lada et al. distribution as read from their Figure 4 and a version from our
data after smoothing the {\it Herschel}-derived column density with an 80\as\ HPW Gaussian to try to duplicate the Lada et al. result.
Both our native-resolution function as well as the smoothed version are well above the Lada et al. function at all values of A$_K$.
It is possible that part of this difference is due to the fact that our observations cover only about half of the area mapped by
\citet{lada09} that has an extinction above A$_K$ = 0.2 mag, though it is difficult to imagine quantitatively how such an areal
difference could have such a large effect on the derived mass function.  Another significant difference, besides just the
basic technique, is the higher angular resolution of our data compared to the NICER optical extinction method, $\sim$ 35\as\ versus $\sim$ 80\as,
but our smoothed mass function is also well above that measured with the NICER technique.  
Yet another possible explanation for the difference is that there may exist a population of dust even colder than that sampled by
the {\it Herschel} observations that could be contributing to the extinction-derived dust masses.  The fact, however, that we have clearly
sampled dust down to T$\sim$ 10 K and that the diffuse dust emission is significantly warmer at temperatures of order 14--15 K argues
against this hypothesis.
We have also investigated whether the difference might be due to differences in assumed dust properties.  The underlying
relationship which would affect Figure \ref{cummsf} is the ratio of far-ir optical depth used to derive our
gas column density to the near-ir dust optical depth measured by \citet{lada09}.  If, for example, the K-band
extinction were smaller at any given value of far-ir optical depth, then the left side (low A$_K$) part of the
{\it Herschel}-derived mass function would shift closer to that found by \citet{lada09}.  On the other hand,
however, the right side of our mass function (high A$_K$) would then drop far below the corresponding part of
the NICER-derived mass function, particularly for the smoothed (80\as) version of our mass function.
Finally, it is also possible that the 2MASS-derived NICER extinctions do not sample well the highest column-density
parts of the cloud.
In any case, it will be very interesting
to compare these results with {\it Herschel}-derived column density maps for the OMC to see if a similar discrepancy exists
in that cloud between the NICER and SED-fitting methods.

The total mass in our mapped area is $\sim$ 4.9 $\times 10^4$ \msun\ with 4.89 $\times 10^4$ \msun\ above
an extinction of A$_K$ = 0.1 mag with the various assumptions mentioned earlier.  As shown in Table \ref{masstable}, 
these mass values are about a factor of two below those found by \citet{lada09} who observed
a significantly larger area, most of which is occupied by relatively low column density material.  So within the uncertainties the total masses are in reasonable agreement.  
Table \ref{masstable} also illustrates numerically the difference in mass distribution in that our estimated mass above $A_K = 1$ mag is three times larger
than that found by \citet{lada09} despite the smaller area covered in our study.

\begin{figure}
\includegraphics[angle=0,totalheight=100mm]{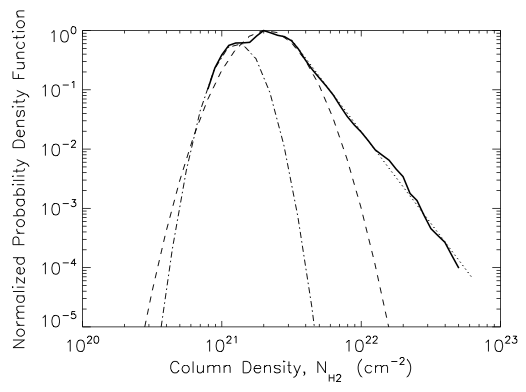}
\figcaption{\label{probdf}
Normalized probability density function of area versus column density.  The dashed line and dash-dot line show two different log-normal
distributions that can fit portions of the PDF.  The dotted line shows a power-law slope of -3.1 that roughly fits the extended
tail of the distribution at high column densities.}
\end{figure}

Figure \ref{probdf} shows the probability density function (PDF) of column density for the AMC.  
This figure also shows three possible fits to portions of the PDF, two log-normal distributions for the central part of the PDF, and a power-law falloff
for the high-extinction end for comparison with other recent studies of Gould Belt clouds.
This observed distribution is qualitatively
similar to other published column density PDFs, but peaks at a moderately lower column density, 2$\times$ 10$^{21}$ cm$^{-2}$, than that for the Aquila region,
4 $\times$ 10$^{21}$ cm$^{-2}$ \citep{andre11}.
The power-law slope of -3  at high extinctions is comparable to that found for Aquila, as well as for the
Rosette Nebula by \citet{schneider12}.

Qualitatively Figure 
\ref{probdf} supports the suggestion by \citet{lada09} that the AMC has relatively less
high-column-density material than more prolific star-forming regions like the Orion Molecular Cloud.  This conclusion will
be able to be further quantified as the {\it Herschel} data on the OMC become available for comparison, since it is clear already
from our results that the resolution of the observations and technique used  may be important to the detailed results.  Further analysis will hopefully
also provide some insight into the underlying physical mechanisms that lead to these differences. 

\subsection{Star Formation Versus Column Density}

\begin{figure}[h]
\includegraphics[angle=0,totalheight=70mm]{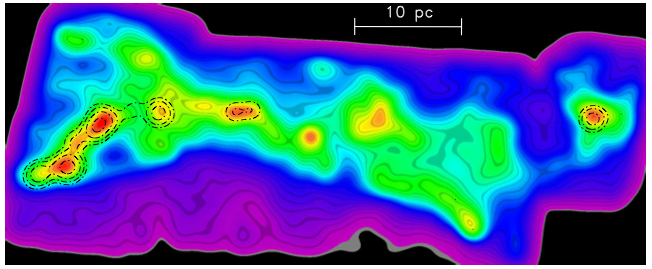}
\figcaption{\label{ysogas}
Derived $N_{H2}$ column density smoothed with a 0.2\degree\ (HPW) Gaussian shown in colored contours versus the
YSO surface density smoothed in the same way in black dash-dot contours.  The YSO contours are in steps of 1, 2, 4, and 8 YSOs per smoothed beam for the Class 0--II objects.
The smoothed column density ranges from 1 to 8.6 $\times 10^{21}$ cm$^{-2}$.}
\end{figure}

We have already discussed the fact that virtually all the young stellar objects are found along the high
column density filamentary structure shown in Figure \ref{tempcold}.  In fact, if we confine our sample to
the Class 0 through II SED objects that are likely to be young enough that they are still close to their
birthplaces in the cloud, there is only one YSO outside the regions of the cloud with N$_{H2} < 5 \times\ 10^{21}$ cm$^{-2}$
as mentioned earlier.
We can investigate whether there is
a quantitative as well as qualitative correlation by smoothing both distributions and comparing the YSO surface density
with the gas surface density.  Figure \ref{ysogas} shows the result of this comparison, where we plot the surface density
of the 68 Class 0--II YSOs
and the {\it Herschel}-derived gas column density, both smoothed with a 0.2\degree (1.6 pc) half-power-width Gaussian.  For the YSOs we used
the union of {\it Spitzer} (H. Broekhoven-Fiene et al. 2013, in preparation) and {\it Herschel} (this paper) objects.
The two highest concentrations of YSOs are found in the LkH$\alpha$101 cluster and in a clump in the northern filament
about 3/4\degree\ north of LkH$\alpha$101.  The derived column density is also highest in these two areas as smoothed 
to a 0.2\degree\ half-power-width.  There is, however, a less perfect correlation between YSO surface density and derived
gas column density at the intermediate levels, but it is certainly true that the greatest number of YSOs are found in the
regions with the highest concentration of dust and gas.  Conversely, in lower column density areas, but still above the general
background level, essentially no YSOs are found.  

\begin{figure}[h]
\includegraphics[angle=0,totalheight=120mm]{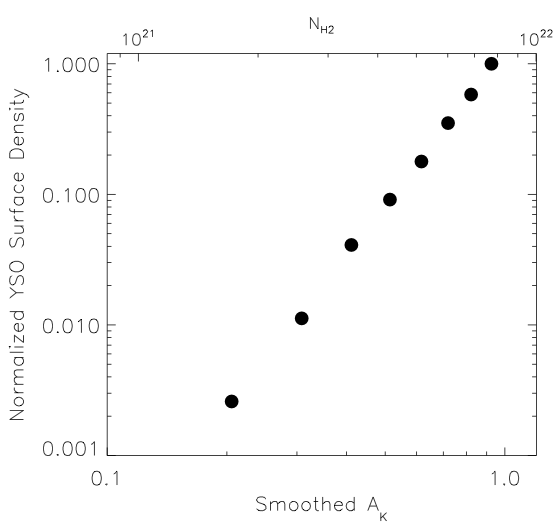}
\figcaption{\label{akvsyso}
Plot of the ratio of smoothed YSO surface density to smoothed column density (normalized to the peak) relative to the
smoothed column density (expressed in K magnitude extinction $A_K$), both from Figure \ref{ysogas}.  }
\end{figure}

\citet{kennevans12} have reviewed this subject extensively in the context of both Galactic and extragalactic star
formation.  Likewise \citet{lada10} and \citet{heid10} have attempted to compare star formation rates and gas surface
densities in multiple local star-forming environments.  All of these studies find a roughly power-law relation
between gas density and star formation rate over some range of gas densities.
We can quantify our own conclusions by computing the ratio of the two maps plotted in Figure \ref{ysogas} as a function of
the derived column density.  Figure \ref{akvsyso} shows the average of this ratio in ten column density bins expressed
as $A_K$ (mag) from the map smoothed with a 0.2\degree\ Gaussian.  This plot shows clearly that there is a strong power-law
relationship between star formation and column density in the AMC.  The surface density of young stellar objects increases rapidly
at the highest column densities.  The slope of this relationship is 4.0. This conclusion does not depend strongly on the details of our sample since the distribution
of {\it Spitzer} and {\it Herschel} YSOs is similar within the small-number statistics, and our derived column density map is
qualitatively similar to that of \citet{dob05}.  This slope of 4.0 is comparable to, but slightly less than the slope of 4.6 derived
by \citet{heid10} for an ensemble of star-forming regions at comparable levels of extinction (surface density).  
We do not, however, see any obvious break point in the relation between YSO surface
density and gas column density, though such a break point might be masked by the smoothing process necessary to
deal with relatively small number of YSOs.
Note also that our maximum
smoothed extinction is A$_K \sim 1.0$ mag, equivalent to a visual extinction of A$_V \sim 8$ mag which is just below the level where
\citet{heid10} see a break in their power-law slope.

\subsection{Ionizing Environment}

Sharp edges suggesting shaping by photoionization are seen along the southern border of the cloud (Fig. \ref{aurcalrgb}), 
particularly in the west near $\ell =160.5\degr , {\it b}=-9.5$\degr, even down to low-column densities. 
The effects of photoionization, however, are modest because there is no evidence for a temperature gradient indicative of dust heating.  
The sharp cloud edges are reminiscent of the similar but stronger effects  seen in the Oph North region (upper Scorpius) 
which is being photoionized and shaped by the runaway O star $\zeta$ Ophiuchi \citep{hat12}.

A possible source of photoionization is the  O7.5 III  star $\xi$~Per ($\ell=160.4\degr, b=-13.1\degr$). The star $\xi$~Per  
illuminates the California Nebula, a bright infrared nebula located between the star and  the AMC in projection.  
\cite{hoo01} model $\xi$~Per as a runaway O star that was ejected from the Per OB2 association, and that now has a distance of 360 pc. 
The proper motion and distance uncertainties, however, are also consistent with the interpretation that  $\xi$~Per is closer to the AMC cloud, 
and hence able to influence the AMC cloud boundary. In any case, the influence of $\xi$~Per on the structure of the AMC is minimal, 
and there is no indication of enhanced star formation along the southwest cloud boundary.






\section{Comparison of the AMC and OMC}\label{compare}

As mentioned above, a comparable {\it Herschel} study of the OMC is not yet complete so it is not possible yet to make a detailed
comparison of the star formation and interstellar medium between the two giant molecular clouds in the far-infrared.
We can, however, discuss briefly the differences in star formation rate on the basis of our deep {\it Herschel}
observations.  H. Broekhoven-Fiene et al. (2013, in preparation) have shown that on the basis of {\it Spitzer} searches for infrared-excess objects,
the AMC appears to have about 5\% of the YSO population over a similar area as that of the OMC \citep{megeath12}.
Our {\it Herschel} photometry has discovered a few additional young objects on the basis of 70 \micron\ and 160 \micron\
fluxes, but there is clearly not a significant population of deeply buried YSOs.  This ratio of 20:1 for the
star formation rates is, though, not as great as the ratio of the incidence of very massive stars.  For example, the seminal
study of \citet{blaauw64} found more than 50 O and early B stars in the Orion OB associations, whereas the AMC
probably has only one early B star.   This suggests that whatever the reasons for the lower star formation rate
in the AMC despite its total mass, the rate for high mass stars is depressed even more in the AMC relative to
that in the OMC.

\section{Summary}\label{summ}

We have completed the census of star formation in the AMC that began with the {\it Spitzer} Gould Belt survey
of H. Broekhoven-Fiene et al. (2013, in preparation). We have found a modest number of additional YSOs, 11, several of which exhibit quite cold SEDs that peak at 150 \micron\--200\micron.
We also mapped a subsection of the {\it Herschel} area with Bolocam at 1.1 mm and found 18 cold dust sources whose fluxes
are well correlated with the dust temperature and column density derived at shorter wavelengths with {\it Herschel}.
We have analyzed the distribution of column density and found a strong non-linear relation between column density and
YSO surface density.  We have compared our derived cumulative mass fraction with that found by \citet{lada09} with
the NICER method and noted some differences that may be due to a combination of factors including:
area covered, angular resolution, and details of the methods.  The cumulative mass fraction and the probability
density function for the column density are both qualitatively similar to other clouds for which they have been
derived but may suggest that the AMC is dominated by lower column density material than other clouds with
higher rates of star formation as suggested by \citet{lada09}.  The star formation rate in the AMC appears to be a
factor of 20 below that in the OMC for typical stars, but an even greater difference  exists at the high-mass end
of the IMF.

\section{Acknowledgments}

We thank Nicholas Chapman and Giles Novak for their help with the CSO/Bolocam observations, and we thank
the anonymous referee who provided a number of comments that noticeably improved this paper.
Support for this work, as part of the NASA Herschel Science Center data analysis funding program, 
was provided by NASA through a contract issued by the Jet Propulsion Laboratory, California 
Institute of Technology to the University
of Texas.
Partial support for T.L.B. was provided by NASA through contract
1433108 issued by the Jet Propulsion Laboratory, California Institute of
Technology, to the Smithsonian Astronomical Observatory.

This publication makes use of data products from the Wide-field Infrared Survey Explorer, which is a joint project of the University of California, Los Angeles, and the Jet Propulsion Laboratory/California Institute of Technology, funded by the National Aeronautics and Space Administration.
This research has also made use of the SIMBAD database, operated at CDS, Strasbourg, France.

\clearpage

\begin{table}[h]
\caption{AOR List \label{obsidtbl}}
\vspace {3mm}
\begin{tabular}{lccc}
\tableline
\tableline
AOR Name  & ObsID  &  Field Center & Comments \cr
\tableline

SPParallel-aurwest-orth & 1342239276   & 04 09 53.0 +39 59 30 & Western End \cr
SPParallel-aurwest-norm & 1342239277   & 04 10 00.0 +40 01 27 & Western End \cr
SPParallel-aurcntr-orth & 1342239278   & 04 18 57.0 +37 45 09 & Central Region \cr
SPParallel-aurcntr-norm & 1342239279   & 04 19 03.5 +37 44 54 & Central Region \cr
PPhoto-secluster-orth  &  1342239441   & 04 30 30.0 +35 30 00 & LkHa101 Cluster \cr
PPhoto-secluster-norm  &  1342239442   & 04 30 30.0 +35 30 00 & LkHa101 Cluster \cr
SPParallel-aureast-orth & 1342240279   & 04 30 20.7 +35 50 57 & Eastern End \cr
SPParallel-aureast-norm & 1342240314   & 04 30 19.9 +37 50 58 & Eastern End \cr

\tableline

\tableline
\end{tabular}
\end{table}

\begin{deluxetable}{lcclcccccccccc}
\setlength{\tabcolsep}{0.04in}
\tabletypesize{\footnotesize}
\rotate
\tablecolumns{14}
\tablecaption{{\it Herschel} Source Fluxes And Derived Quantities  \label{fluxtbl}}
\tablewidth{0pt}
\tablehead{
\colhead{Src} &
\colhead{YSO\tablenotemark{a}} &
\colhead{SIMBAD} &
\colhead{RA/Dec Center (J2000)} &
\colhead{MIR} &
\colhead{$\alpha$} &
\colhead{YSO} &
\colhead{L$_{bol}$} &
\colhead{T$_{bol}$} &
\colhead{F$_\nu$ 22/24 \micron} &
\colhead{F$_\nu$ 70 \micron\tablenotemark{b}} &
\colhead{F$_\nu$ 70 \micron} &
\colhead{F$_\nu$ 160 \micron} &
\colhead{F$_\nu$ 160 \micron} \\
\colhead{} &
\colhead{} &
\colhead{Type} &
\colhead{h\quad m\quad s\quad \quad \degree\quad \am\quad \as\quad} &
\colhead{} &
\colhead{} &
\colhead{Class} &
\colhead{\lsun} &
\colhead{K} &
\colhead{mJy} &
\colhead{(PSF) mJy} &
\colhead{(Aper) mJy} &
\colhead{(PSF) mJy} &
\colhead{(Aper) mJy} 
}
\startdata
  1 &  Y &   IR  & 04 09 02.16 $+$40 19 11.4 & WISE &  0.56 &    I &  2.16 &  152 &  980$\pm$  91 & 3880$\pm$   98 & 3950$\pm$  120 & 3360$\pm$  290 & 4600$\pm$  110\\
  2 &    &       & 04 09 54.71 $+$40 06 39.9 & SpGB &  0.99 &    I &  0.11 &   71 & 7.16$\pm$0.70 &  144$\pm$ 7.6 &  185$\pm$  27 &  818$\pm$  92 & 2210$\pm$   48\\
  3 &  Y &       & 04 10 02.81 $+$40 02 43.9 & SpGB &  2.37 &    0 &  0.37 &   31 & 2.08$\pm$0.68 &  647$\pm$  26 &  583$\pm$  36 & 3370$\pm$  750 & 12600$\pm$    92\\
  4 &    &       & 04 10 04.53 $+$40 02 37.5 & SpGB &  1.78 &    0 &  0.29 &   28 & $<$  2.00 &  221$\pm$  12 &  348$\pm$ 120 & 3430$\pm$  690 & 18800$\pm$   140\\
  5 &  Y &       & 04 10 05.88 $+$40 02 37.0 & SpGB &  1.04 &    0 &  0.61 &   42 & 3.51$\pm$0.68 &  592$\pm$  46 &  989$\pm$ 590 & 6660$\pm$ 1800 & 31800$\pm$   280\\
  6 &    &   IR  & 04 10 07.08 $+$40 02 34.6 & WISE &  1.19 &    0 &  2.00 &   48 &  156$\pm$ 4.5 & 2230$\pm$  110 & 4080$\pm$ 3000 & 18600$\pm$  2700 & 54000$\pm$   550\\
  7 &  Y &   IR  & 04 10 08.58 $+$40 02 23.2 & SpGB &  1.05 &    I & 12.11 &   97 & 4770$\pm$  470 & 25400$\pm$   390 & 29000$\pm$   660 & 33800$\pm$  4400 & 58600$\pm$   670\\
  8 &  Y &       & 04 10 11.29 $+$40 01 24.4 & SpGB &  1.94 &    0 &  0.55 &   47 & 41.1$\pm$ 3.8 & 1610$\pm$   25 & 1550$\pm$   59 & 2670$\pm$  190 & 3850$\pm$   89\\
  9 &  Y &  Em*  & 04 10 40.95 $+$38 07 52.4 & WISE &  0.32 &    I & 40.53 &  239 & 15001$\pm$   170 & 50100$\pm$  1000 & 52800$\pm$  1400 & 53000$\pm$  4200 & 66600$\pm$  1200\\
 10 &  Y &  Em*  & 04 10 49.03 $+$38 04 43.8 & SpGB & -0.25 &    F &  0.47 &  337 &  123$\pm$  11 &  240$\pm$  11 &  234$\pm$  29 &  923$\pm$  61 & 1860$\pm$   47\\
 11 &    &       & 04 12 40.54 $+$38 14 26.8 & SpGB &  0.96 &  I/0 &  0.02 &   65 & 0.83$\pm$0.20 & 82.9$\pm$ 9.0 & \nodata & 81.4$\pm$  23 & \nodata\\
 12 &  Y &  RNe  & 04 21 37.77 $+$37 34 41.1 & SpGB &  0.04 &    F &  3.40 &  290 &  223$\pm$  21 & 3450$\pm$   94 & 4620$\pm$  320 & 13900$\pm$   930 & 25900$\pm$   250\\
 13 &  Y &       & 04 21 40.58 $+$37 33 58.3 & SpGB &  0.94 &    I &  0.56 &  123 &  241$\pm$  23 &  910$\pm$  27 &  906$\pm$  44 & 1450$\pm$  290 & 4170$\pm$   62\\
 14 &    &       & 04 24 59.04 $+$37 17 52.9 & SpGB &  0.84 &    I &  0.04 &   93 & 7.63$\pm$0.73 &  138$\pm$  11 &  117$\pm$  30 & 98.9$\pm$  38 &  112$\pm$  41\\
 15 &    &   IR  & 04 25 07.83 $+$37 15 19.3 & SpGB &  2.45 &    0 &  0.94 &   36 & 6.85$\pm$0.67 & 2670$\pm$   48 & 2700$\pm$   76 & 6360$\pm$  620 & 8200$\pm$  210\\
 16 &  Y &   IR  & 04 25 38.30 $+$37 06 59.2 & SpGB &  1.36 &  I/0 &  0.57 &   55 & 59.1$\pm$ 5.5 & 1160$\pm$   29 & 1090$\pm$  110 & 3490$\pm$  400 & 9330$\pm$   88\\
 17 &  Y &   IR  & 04 25 39.60 $+$37 07 06.5 & SpGB & -0.61 &   II &  2.95 &  459 &  727$\pm$  68 &  630$\pm$  15 &  637$\pm$ 190 & 1350$\pm$  340 & 10100$\pm$    89\\
 18 &  Y &       & 04 28 14.90 $+$36 30 27.4 & SpGB & -0.27 &    F &  0.13 &  296 & 61.3$\pm$ 5.7 &  109$\pm$  14 & \nodata & 99.0$\pm$  41 &  792$\pm$  54\\
 19 &  Y &       & 04 28 35.07 $+$36 25 05.2 & SpGB &  0.49 &    I &  0.91 &  158 &  237$\pm$  22 & 1660$\pm$   43 & 1590$\pm$   57 & 2400$\pm$  200 & 3590$\pm$   69\\
 20 &  Y &       & 04 28 37.87 $+$36 24 54.9 & SpGB & -0.23 &    F &  1.20 &  338 &  204$\pm$  19 & 1290$\pm$   31 & 1230$\pm$   53 & 1950$\pm$  180 & 3620$\pm$   65\\
 21 &  Y &   IR  & 04 28 38.54 $+$36 25 28.1 & SpGB &  0.94 &    I &  0.34 &   83 &  143$\pm$  13 &  723$\pm$  21 &  704$\pm$  38 & 1010$\pm$  200 & 2750$\pm$   56\\
 22 &  Y &       & 04 28 43.66 $+$36 28 37.6 & SpGB &  1.16 &  I/0 &  0.93 &   68 &  192$\pm$  18 & 2630$\pm$   73 & 2670$\pm$   90 & 3150$\pm$  290 & 4960$\pm$   95\\
 23 &  Y &   IR  & 04 28 55.24 $+$36 31 21.6 & SpGB &  0.60 &    I &  2.33 &  138 &  752$\pm$  70 & 4490$\pm$  140 & 4560$\pm$  150 & 5870$\pm$  720 & 9080$\pm$  210\\
 24 &  Y &   IR  & 04 29 54.15 $+$36 11 56.3 & WISE & -0.40 &   II &  1.00 &  322 &  534$\pm$  50 &  726$\pm$  18 &  657$\pm$  36 &  712$\pm$  55 &  960$\pm$  45\\
 25 &  Y &  Y*O  & 04 29 55.05 $+$35 18 04.8 & SpGB & -0.22 &    F &  0.43 &  268 &  135$\pm$  13 &  669$\pm$  77 & 1480$\pm$  170 & \nodata & 4760$\pm$  100\\
 26 &  Y &       & 04 29 59.31 $+$36 10 17.5 & WISE & -0.34 &   II &  0.12 &  434 & 10.6$\pm$ 1.0 & 88.8$\pm$  11 & 58.0$\pm$  34 &  207$\pm$  51 &  689$\pm$  42\\
 27 &  Y &       & 04 30 14.89 $+$36 00 08.3 & SpGB &  0.96 &    I &  0.12 &  120 & 48.2$\pm$ 4.5 &  198$\pm$ 10 &  163$\pm$  28 &  297$\pm$  33 &  490$\pm$  39\\
 28 &  Y &  PN?  & 04 30 15.68 $+$35 56 57.8 & WISE & -0.46 &   II & 19.21 &  368 & 8160$\pm$  130 & 14900$\pm$   420 & 15500$\pm$   420 & 5510$\pm$  900 & 10500$\pm$   170\\
 29 &  Y &       & 04 30 24.58 $+$35 45 20.8 & SpGB &  0.51 &    I &  2.88 &  150 & 1400$\pm$  130 & 4840$\pm$  100 & 4960$\pm$  130 & 4830$\pm$  210 & 5480$\pm$  120\\
 30 &  Y &       & 04 30 26.91 $+$35 45 51.9 & SpGB &  0.39 &    I &  0.08 &  117 & 14.4$\pm$ 1.4 &  179$\pm$ 7.2 &  173$\pm$  37 &  243$\pm$  57 & 1110$\pm$   49\\
 31 &  Y &  Y*O  & 04 30 27.59 $+$35 09 17.5 & SpGB &  1.04 &    I &  9.26 &  105 & 1580$\pm$  160 & 22500$\pm$   840 & 28100$\pm$   520 & 32400$\pm$  3000 & 45400$\pm$   670\\
 32 &  Y &       & 04 30 27.71 $+$35 46 14.6 & SpGB & -0.31 &   II &  0.23 &  331 & 95.0$\pm$ 8.8 &  120$\pm$ 8.7 & 97.0$\pm$  33 &  275$\pm$  93 & 1220$\pm$   45\\
 33 &  Y &       & 04 30 28.50 $+$35 47 44.5 & SpGB & -0.36 &   II &  0.18 &  395 & 33.4$\pm$ 3.1 & 96.6$\pm$ 9.9 & 70.7$\pm$  29 &  262$\pm$  39 &  624$\pm$  40\\
 34 &  Y &       & 04 30 30.05 $+$35 06 39.9 & SpGB & -0.44 &   II &  0.13 &  356 & 38.5$\pm$ 3.7 &  117$\pm$  11 &  162$\pm$  35 &  133$\pm$  77 &  627$\pm$  41\\
 35 &  Y &       & 04 30 30.50 $+$35 51 44.1 & SpGB &  0.45 &    I &  0.42 &  152 &  187$\pm$  17 &  590$\pm$  13 &  538$\pm$  32 & 1050$\pm$   77 & 1530$\pm$   50\\
 36 &  Y &       & 04 30 31.53 $+$35 45 14.3 & SpGB & -0.36 &   II &  1.10 &  352 &  403$\pm$  37 &  595$\pm$  16 &  581$\pm$  35 &  931$\pm$ 120 & 1830$\pm$   54\\
 37 &  Y &   IR  & 04 30 32.32 $+$35 36 13.4 & SpGB &  0.07 &    F &  0.72 &  149 &  292$\pm$  27 & 1440$\pm$   35 & 1250$\pm$   57 & 1470$\pm$   84 & 1720$\pm$   52\\
 38 &  Y &       & 04 30 36.74 $+$35 54 36.8 & SpGB &  1.11 &    I &  2.67 &   92 &  529$\pm$  50 & 6280$\pm$  160 & 6390$\pm$  190 & 9750$\pm$  870 & 13400$\pm$   250\\
 39 &  Y &  TT?  & 04 30 37.42 $+$35 50 31.4 & SpGB & -0.10 &    F &  1.65 &  275 &  390$\pm$  36 & 1720$\pm$   33 & 1930$\pm$  220 & 4140$\pm$  600 & 13100$\pm$   170\\
 40 &  Y &       & 04 30 37.81 $+$35 51 01.2 & SpGB &  2.10 &    0 &  0.72 &   38 & 9.82$\pm$0.92 & 2040$\pm$   51 & 1940$\pm$   83 & 4560$\pm$  770 & 7070$\pm$  160\\
 41 &  Y &       & 04 30 38.14 $+$35 49 59.5 & SpGB &  2.15 &    0 &  0.89 &   44 & 70.5$\pm$ 6.5 & 2190$\pm$   47 & 2180$\pm$  100 & 5270$\pm$  550 & 10500$\pm$   160\\
 42 &    &       & 04 30 38.38 $+$35 50 22.6 & SpGB &  2.07 &    0 &  0.58 &   32 & $<$  3.00 & 1050$\pm$   21 &  944$\pm$ 370 & 5210$\pm$ 1800 & 14500$\pm$   180\\
 43 &  Y &   IR  & 04 30 38.76 $+$35 54 40.2 & SpGB &  0.21 &    F &  0.23 &  201 & 53.4$\pm$ 4.9 &  263$\pm$  12 & \nodata &  822$\pm$ 220 & 6870$\pm$  160\\
 44 &  Y &   IR  & 04 30 39.18 $+$35 52 02.1 & SpGB & -0.09 &    F &  2.41 &  249 &  899$\pm$  85 & 2510$\pm$   71 & 2920$\pm$   85 & 5070$\pm$  600 & 9480$\pm$  110\\
 45 &  Y &       & 04 30 41.13 $+$35 29 40.5 & SpGB &  1.18 &    I &  0.93 &   81 &  176$\pm$  16 & 2230$\pm$   57 & 2310$\pm$   70 & 3620$\pm$  240 & 6480$\pm$   98\\
 46 &  Y &  PN?  & 04 30 44.13 $+$35 59 50.8 & SpGB &  0.24 &    F &  2.81 &  254 & 1270$\pm$  120 & 2390$\pm$   52 & 2170$\pm$   73 & 3120$\pm$  200 & 3860$\pm$   88\\
 47 &  Y &       & 04 30 46.20 $+$34 58 55.6 & SpGB &  1.57 &    0 &  0.31 &   49 & 26.9$\pm$ 2.5 &  719$\pm$  20 &  826$\pm$ 130 & 1840$\pm$  240 & 10900$\pm$   240\\
 48 &  Y &       & 04 30 47.24 $+$35 07 42.6 & SpGB & -0.49 &   II &  0.16 &  379 & 51.8$\pm$ 4.8 &  107$\pm$  13 &  114$\pm$  34 & 84.1$\pm$  56 &  345$\pm$  42\\
 49 &    &  PN?  & 04 30 47.90 $+$34 58 37.3 & SpGB &  1.89 &    0 &  1.27 &   33 & 9.09$\pm$ 1.6 & 2400$\pm$   49 & 2430$\pm$  110 & 11200$\pm$  1300 & 18700$\pm$   340\\
 50 &  Y &   IR  & 04 30 48.42 $+$35 37 54.4 & SpGB &  1.01 &    I &  1.92 &   92 &  452$\pm$  42 & 5020$\pm$  140 & 5160$\pm$  150 & 5620$\pm$  580 & 8220$\pm$  170\\
 51 &  Y &  PN?  & 04 30 48.54 $+$34 58 52.7 & SpGB & -0.20 &    F &  0.79 &  236 &  677$\pm$  63 &  466$\pm$  26 &  618$\pm$ 370 &  623$\pm$ 450 & 15900$\pm$   320\\
 52 &  Y &       & 04 30 49.18 $+$34 56 10.2 & SpGB & -0.02 &    F &  0.48 &  236 &  277$\pm$  26 &  475$\pm$  15 &  453$\pm$  34 &  382$\pm$  25 &  227$\pm$  41\\
 53 &  Y &  Or*  & 04 30 49.63 $+$34 57 27.9 & SpGB & -0.69 &   II &  2.54 &  451 &  677$\pm$  63 & 1210$\pm$   42 & 1230$\pm$   52 & 1240$\pm$  220 & 1720$\pm$   54\\
 54 &  Y &       & 04 30 52.00 $+$34 50 08.4 & WISE & -0.26 &    F &  0.31 &  322 &  114$\pm$  11 &  253$\pm$ 9.9 &  316$\pm$  33 &  364$\pm$  50 &  223$\pm$  44\\
 55 &  Y &       & 04 30 53.41 $+$34 56 26.4 & SpGB &  0.54 &    I &  0.06 &  130 & 27.2$\pm$ 2.5 & 81.3$\pm$ 9.6 & 74.2$\pm$  35 &  160$\pm$  47 &  607$\pm$  43\\
 56 &  Y &       & 04 30 55.92 $+$34 56 48.3 & SpGB &  0.54 &    I &  0.25 &  134 &  141$\pm$  13 &  377$\pm$  12 &  385$\pm$  38 &  503$\pm$  59 & 1040$\pm$   46\\
 57 &  Y &   IR  & 04 30 56.49 $+$35 30 04.5 & SpGB &  1.54 &    I &  1.00 &   70 &  302$\pm$  28 & 2560$\pm$   86 & 2670$\pm$   87 & 3180$\pm$  460 & 5150$\pm$  110\\
 58 &    &       & 04 30 57.19 $+$34 53 53.6 & WISE &  0.38 &    I &  0.11 &  175 & 39.9$\pm$ 3.7 &  129$\pm$ 10 &  133$\pm$  30 &  353$\pm$  53 & 1130$\pm$   43\\
 59 &    &       & 04 31 14.67 $+$35 56 50.6 & SpGB &  0.76 &    I &  0.09 &  103 & 26.4$\pm$ 2.5 &  178$\pm$10 &  150$\pm$  35 &  261$\pm$  45 &  590$\pm$  46\\
 60 &    &       & 04 34 53.15 $+$36 23 27.9 & WISE & -0.44 &   II &  2.76 &  367 & 1089$\pm$   15 & 1840$\pm$   46 & 1810$\pm$   61 & 1770$\pm$  190 & 3140$\pm$   77\\

\tablebreak
\enddata
\tablenotetext{a}{Identified as YSO candidate by H. Broekhoven-Fiene et al. (2013, in preparation)  }
\tablenotetext{b}{Absolute calibration uncertainty estimated as $\pm$15\% for all {\it Herschel} photometry}

\end{deluxetable}

\begin{deluxetable}{clcccccc}
\tabletypesize{\small}
\rotate
\tablecolumns{7}
\tablecaption{1.1 mm Source Fluxes  \label{mmfluxtbl}}
\tablewidth{0pt}
\tablehead{
\colhead{CSO} &
\colhead{RA/Dec Center (J2000)} &
\colhead{{\it Herschel}} &
\colhead{F$_\nu$ Fit} &
\colhead{F$_\nu$ 40\as} &
\colhead{F$_\nu$ 80\as} &
\colhead{F$_\nu$ 120\as}  \\
\colhead{Src\#} &
\colhead{h\quad m\quad s\quad \quad \degree\quad \am\quad \as\quad\quad} &
\colhead{Sources} &
\colhead{Jy} &
\colhead{Jy} &
\colhead{Jy} &
\colhead{Jy} 
}
\startdata
SSTGB & 04 01 24.5 $+$41 01 49 &  & 0.63$\pm$0.07 & 0.26$\pm$0.029 & 0.53$\pm$0.058 & 0.66$\pm$0.083\\
  1 & 04 10 41.4 $+$38 07 59 &    9 & 0.71$\pm$0.22 & 0.48$\pm$0.13 & 0.72$\pm$0.25 & 0.63$\pm$0.38\\
  2 & 04 18 21.3 $+$38 01 36 & 3C111  & 1.98$\pm$0.24 & 1.25$\pm$0.15 & 2.03$\pm$0.29 & 2.33$\pm$0.41\\
  3 & 04 19 27.6 $+$38 00 03 &  & 0.53$\pm$0.23 & 0.18$\pm$0.12 & 0.60$\pm$0.25 & 0.70$\pm$0.36\\
  4 & 04 19 29.4 $+$37 59 43 &  & 0.57$\pm$0.22 & 0.21$\pm$0.12 & 0.65$\pm$0.24 & 0.79$\pm$0.36\\
  5 & 04 21 17.4 $+$37 33 16 &  & 0.53$\pm$0.21 & 0.12$\pm$0.09 & 0.42$\pm$0.18 & 0.78$\pm$0.28\\
  6 & 04 21 38.5 $+$37 33 54 &   13 & 0.87$\pm$0.33 & 0.20$\pm$0.10 & 0.59$\pm$0.20 & 0.79$\pm$0.29\\
  7 & 04 25 38.7 $+$37 07 08 &   16  17 & 2.30$\pm$0.48 & 0.27$\pm$0.13 & 1.11$\pm$0.27 & 1.95$\pm$0.41\\
  8 & 04 28 37.5 $+$36 25 27 &   19  20  21 & 1.01$\pm$0.30 & 0.20$\pm$0.13 & 0.80$\pm$0.25 & 1.27$\pm$0.38\\
  9 & 04 30 16.0 $+$35 16 57 & LkH$\alpha$101 & 7.42$\pm$0.72 & 0.59$\pm$0.16 & 2.25$\pm$0.33 & 3.94$\pm$0.51\\
 10 & 04 30 25.6 $+$35 15 07 &  & 2.85$\pm$0.51 & 0.14$\pm$0.13 & 0.83$\pm$0.26 & 1.89$\pm$0.40\\
 11 & 04 30 28.4 $+$35 09 27 &   31 & 0.55$\pm$0.20 & 0.28$\pm$0.11 & 0.58$\pm$0.22 & 0.79$\pm$0.33\\
 12 & 04 30 31.3 $+$35 44 49 &   36 & 0.71$\pm$0.27 & 0.12$\pm$0.11 & 0.42$\pm$0.21 & 0.93$\pm$0.33\\
 13 & 04 30 37.6 $+$35 54 36 &   38  43 & 0.93$\pm$0.30 & 0.37$\pm$0.13 & 0.87$\pm$0.27 & 0.78$\pm$0.40\\
 14 & 04 30 39.2 $+$35 50 22 &   39  40  41  42 & 0.55$\pm$0.22 & 0.28$\pm$0.12 & 0.67$\pm$0.25 & 0.70$\pm$0.37\\
 15 & 04 30 40.8 $+$35 29 04 &   45 & 0.62$\pm$0.23 & 0.26$\pm$0.13 & 0.72$\pm$0.26 & 1.27$\pm$0.39\\
 16 & 04 30 41.4 $+$35 29 58 &   45 & 1.15$\pm$0.31 & 0.31$\pm$0.13 & 1.02$\pm$0.27 & 1.36$\pm$0.40\\
 17 & 04 30 47.8 $+$35 37 26 &   50 & 0.49$\pm$0.22 & 0.20$\pm$0.12 & 0.55$\pm$0.25 & 0.89$\pm$0.37\\
 18 & 04 30 48.5 $+$34 58 37 &   47  49  51 & 0.85$\pm$0.20 & 0.49$\pm$0.11 & 0.89$\pm$0.23 & 1.02$\pm$0.33\\

\enddata
\end{deluxetable}

\begin{table}[h]
\caption{Cumulative Mass Versus Extinction \label{masstable}}
\vspace {3mm}
\begin{tabular}{lcc}
\tableline
\tableline
$A_K$ (mag)  & Mass (this study\tablenotemark{a})  &  Mass (Lada09\tablenotemark{b})  \cr
Mag  & \msun & \msun \cr
\tableline

0.0 & 4.9 $\times 10^4$ &  NA \cr
0.1 & 4.89 $\times 10^4$ & 1.12 $\times 10^5$ \cr
0.2 & 4.28 $\times 10^4$ & 5.34 $\times 10^4$ \cr
1.0 & 3.29 $\times 10^3$ & 1.09 $\times 10^3$ \cr
\tableline

\tablenotetext{a}{Total survey area is 16.5 square degrees. }
\tablenotetext{b}{Total survey area is $\sim$80 square degrees. }
\end{tabular}
\end{table}


\clearpage

\clearpage

\end{document}